\documentclass[11pt,twoside,a4paper]{article}
\usepackage{graphicx}
\usepackage[dvips]{epsfig}
\usepackage{amssymb}
\usepackage{rotating}
\usepackage{lscape}
\usepackage{a4wide}
\usepackage[usenames]{color}

\def\gsim{\mathrel{\rlap{\lower4pt\hbox{\hskip1pt$\sim$}}\raise1pt\hbox{$>$}}}

\def\Journal#1#2#3#4{{#1} {\bf #2}, #3 (#4).}
\def\IJMPA{Int. J. Mod. Phys. A}
\def\AP{Ann. Phys.}
\def\CTP{Commun. Theor. Phys.}
\def\EPJA{Eur. Phys. J. A}
\def\JP{J. Phys.}

\def\NIMA{{Nucl. Instrum. Methods} A}
\def\NIMB{{Nucl. Instrum. Methods} B}
\def\NPB{Nucl. Phys. B}

\def\PL{Phys. Lett.}
\def\PLB{Phys. Lett. B}
\def\PR{Phys. Rep.}
\def\PRL{Phys. Rev. Lett.}
\def\PRD{Phys. Rev. D}
\def\PRE{Phys. Rev. E}
\def\PRC{Phys. Rev. C}
\def\PRSTAB{Phys. Rev. ST-AB}
\def\RSI{Rev. Sci. Instrum.}
\def\RPP{Rep. Prog. Phys.}

\def\EPJC{Eur. Phys. J}

\newcommand{\be}{\begin{equation}}
\newcommand{\ee}{\end{equation}}
\newcommand{\ba}{\begin{eqnarray}}
\newcommand{\ea}{\end{eqnarray}}

\def\gsim{\mathrel{\rlap{\lower4pt\hbox{\hskip1pt$\sim$}}\raise1pt\hbox{$>$}}}
\def\lsim{\mathrel{\rlap{\lower4pt\hbox{\hskip1pt$\sim$}}\raise1pt\hbox{$<$}}}
\def\ket#1{\left|\,#1\,\right>}

\def\gsim{\mathrel{\rlap{\lower4pt\hbox{\hskip1pt$\sim$}}\raise1pt\hbox{$>$}}}

\begin{document}
\setcounter{secnumdepth}{4}
\setcounter{tocdepth}{4}

\vfill
\begin{center} \LARGE Proposal  \end{center}
\thispagestyle{empty}
\vfill
\begin{center}
  {\noindent\Huge Measurement of the Spin--Dependence of the
  $\bar{p}p$ Interaction at the AD--Ring}
\end{center}
\begin{center} \LARGE (${\cal PAX}$ Collaboration) \end{center}
\vfill
\begin{center} {\noindent\Large J\"ulich, \today} \end{center}
\vfill
\begin{center} {\noindent\Large submitted to the SPS Committee at CERN} \end{center}
\cleardoublepage

\begin{center} \LARGE   Proposal  \end{center}
\begin{center}
  {\noindent\Huge Measurement of the Spin--Dependence of the
  $\bar{p}p$ Interaction at the AD--Ring}
\end{center}
\begin{center} \LARGE (${\cal PAX}$ Collaboration) \end{center}

\begin{abstract}
We propose to use an internal polarized hydrogen storage cell gas target in the AD ring to determine for the first time the two total spin--dependent $\bar{p}p$ cross sections $\sigma_1$ and $\sigma_2$ at antiproton beam energies in the range from 50 to 450 MeV. 

The data obtained are of interest by themselves for the general theory of $\bar{p}p$ interactions since they will provide a first experimental constraint of the spin--spin dependence of the nucleon--antinucleon potential in the energy range of interest. In addition, measurements of the  polarization buildup of stored antiprotons are required to define the optimum parameters of a future, dedicated Antiproton Polarizer Ring (APR), intended to feed a double--polarized asymmetric $\bar{p}p$ collider with polarized antiprotons. Such a machine has recently been proposed by the PAX collaboration for the new Facility for Antiproton and Ion Research (FAIR) at GSI in Darmstadt, Germany. The availability of an intense stored beam of polarized antiprotons will provide access to a wealth of single-- and double--spin observables, thereby opening a new window on QCD spin physics.

A recent experiment at COSY revealed that $e\vec{p}$ spin--flip cross sections are too small to cause a detectable depolarization of a stored proton beam. This measurement rules out a proposal to use polarized positrons to polarize an antiproton beam by $\overrightarrow{e^+}\bar{p}$ spin--flip interactions. Thus, our approach to provide a beam of polarized antiprotons is based on spin filtering, using an internal polarized hydrogen gas target -- a method that has been tested with stored protons. We expect to produce a polarized antiproton beam with at least ten orders of magnitude higher intensity than a secondary polarized antiproton beam previously available.

Provided that antiproton beams with a polarization of about 15\% can be obtained with the APR, the antiproton machine at FAIR (the High Energy Storage Ring) could be converted into a double--polarized asymmetric $\bar{p}p$ collider by installation of an additional COSY--like ring. In this setup, antiprotons of 3.5 GeV/c collide with protons of 15 GeV/c at c.m. energies of $\sqrt{s} \approx \sqrt{200}$~GeV with a luminosity in excess of $10^{31}$~cm$^{-2}$s$^{-1}$.

The PAX physics program proposed for FAIR has been highly rated, and would include, most importantly, a first direct measurement of the transversity distribution of the valence quarks in the proton, and a first measurement of the moduli and the relative phase of the time--like electric and magnetic form factors $G_\mathrm{E,M}$  of the proton.
\end{abstract}
\begin{center}
\section*{Spokespersons:} 
\end{center}
\begin{center}
Paolo Lenisa\\
Istituto Nazionale di Fisica Nucleare, Ferrara, Italy \\
E--Mail: {\tt lenisa@fe.infn.it}\\ 
\end{center}

\begin{center}
Frank Rathmann\\ 
Institut f\"ur Kernphysik, J\"ulich Center for Hadron Physics, \\Forschungszentrum J\"ulich, Germany \\
E--Mail: {\tt f.rathmann@fz-juelich.de}
\end{center}

\cleardoublepage
\section*{PAX collaboration} 

\begin{center}
\small  C. Barschel, U. Bechstedt, J. Dietrich, N. Dolfus, R. Engels, R. Gebel, H. Hadamek, J. Haidenbauer, C. Hanhart, A. Kacharava, G. Krol, M. K\"uven, G. Langenberg, A. Lehrach, B. Lorentz, R. Maier, S.~Martin, U.-G.~Mei{\ss}ner, M.~Nekipelov, N.N. Nikolaev, D. Oellers, G. d'Orsaneo, D. Prasuhn, F.~Rathmann, M. Retzlaff, J.~Sarkadi, R.~Schleichert, H. Seyfarth, A. Sibirtsev, D. Sp\"olgen, H.J.~Stein, H. Stockhorst, H.~Str\"oher, Chr.~Weidemann, D. Welsch, and P. Wieder\\
{\small \it Institut f\"ur Kernphysik, J\"ulich Center for Hadron Physics, Forschungszentrum   J\"ulich GmbH, 52425 J\"ulich, Germany}
 \end{center}

\begin{center}
\small L. Barion, S. Bertelli, V. Carassiti, G. Ciullo, M. Contalbrigo,  A. Cotta--Ramusino, P.F. Dalpiaz, A.~Drago, G. Guidoboni, P. Lenisa, L. Pappalardo, G. Stancari, M. Stancari, and M. Statera \\
{\small\it Universita di Ferrara and INFN, 44100 Ferrara, Italy}
\end{center}

\begin{center}
\small T. Azarian, A. Kulikov, V. Kurbatov, G. Macharashvili, S. Merzliakov, I.N. Meshkov, A. Smirnov, D. Tsirkov, and Yu. Uzikov\\
{\small\it Laboratory of Nuclear Problems, Joint Institute for Nuclear Research, 141980 Dubna, Russia}
\end{center}

\begin{center}
\small S. Barsov, S. Belostotski, K. Grigoryev, P. Kravtsov, M. Mikirtychiants, S. Mikirtychiants, and~A.~Vasilyev\\
{\small\it St. Petersburg Nuclear Physics Institute, 188350 Gatchina, Russia} 
\end{center}
 
\begin{center}
\small F.M. Esser, R. Greven, G. Hansen,  F. Jadgfeld, F. Klehr, H. Soltner, and H. Straatmann\\
{\small \it Zentralabteilung Technologie, J\"ulich Center for Hadron Physics, Forschungszentrum J\"ulich GmbH, 52425 J\"ulich, Germany}
\end{center}

\begin{center}
\small D. Chiladze, A. Garishvili, N. Lomidze, D. Mchedlishvili, M. Nioradze, and M. Tabidze\\
{\small\it High Energy Physics Institute, Tbilisi State University, 0186 Tbilisi, Georgia} 
\end{center}

\begin{center}
\small N. Akopov, A. Avetisyan, G. Elbakyan, H. Marukyan, and S. Taroian\\
{\small\it Yerevan Physics Institute, 0036, Yerevan, Armenia
} 
\end{center}

\begin{center}
\small   P. Benati, W. Erven, F.--J. Kayser, H. Kleines, and P. W\"ustner\\
{\small\it Zentralinstitut f\"ur Elektronik, J\"ulich Center for Hadron Physics, Forschungszentrum J\"ulich GmbH, 52425 J\"ulich, Germany} 
\end{center}

\begin{center}
\small   D. Bruncko, J. Ferencei, J. Mu\v{s}insk\'{y}, and J. Urb\'{a}n\\
{\small\it Institute of Experimental Physics, Slovak Academy of Sciences and P.J. \v{S}af\'{a}rik University, Faculty of Science, 040 01 Ko\v{s}ice, Slovakia 
} 
\end{center}

\begin{center}
\small W. Augustyniak, B. Marianski, A. Trzcinski, and P. Zupranski\\
{\it\small Department of Nuclear Reactions, Andrzej Soltan, Institute for Nuclear Studies, 00-681, Warsaw, Poland} 
\end{center}

\begin{center}
\small S. Dymov, A. Nass, and E. Steffens\\
{\small\it Physikalisches Institut II, Universit\"at Erlangen~N\"urnberg, 91058 Erlangen, Germany} 
\end{center}

\begin{center}
\small K. Rathsman, P.--E. Tegné\'{e}r, and P. Th\"orngren Engblom\\
{\small\it Physics Department, Stockholm University, SE-106 91 Stockholm, Sweden} 
\end{center}

\begin{center}
\small   R. De Leo, and G. Tagliente\\
{\small\it INFN, Sez. Bari, 70126 Bari, Italy}
\end{center}

\begin{center}
\small B. K\"ampfer, and S. Trusov \\
{\it\small Institut f\"ur Kern- und Hadronenphysik, Forschungszentrum Rossendorf, 01314 Dresden, Germany; 
Skobeltsyn Institute of Nuclear Physics, Lomonosov Moscow State University, 119991Moscow, Russia }
\end{center}

\begin{center}
\small N. Buttimore\\
{\small\it School of Mathematics, Trinity College, University of Dublin, Dublin 2, Ireland
} 
\end{center}

\begin{center}
\small H.O. Meyer\\
{\small\it Physics Department, Indiana University, Bloomington, IN 47405, USA} 
\end{center}

\cleardoublepage

\tableofcontents

\cleardoublepage
\section{Introduction\label{introduction}}
\pagestyle{myheadings} \markboth{Spin--Dependence
of the $\bar{p}p$ Interaction at the AD}{Introduction}
In this proposal the PAX collaboration lays out a plan to measure the polarization buildup in an antiproton beam at the AD--ring at CERN by filtering with a polarized internal hydrogen target. The study covers energies in the range from 50--450 MeV. The proposed staging for the implementation of the experimental setup shall provide a smooth   operation of the AD ring for  experiments using extracted beams.

The scientific objectives of this experiment are twofold. Firstly, a measurement of the polarization buildup time yields values for the antiproton--proton spin--dependent total cross sections. Apart from the obvious interest for the general theory of $\bar{p}p$ interactions, the know\-ledge of these cross sections is also necessary for the interpretation of unexpected features of the $p\bar{p}$, and other antibaryon--baryon pairs, contained in final states in $J/\Psi$ and $B$--decays. Secondly, an empirical verification of polarization buildup in a stored antiproton beam would pave the way to a future high--luminosity double--polarized antiproton--proton collider, which would provide a unique laboratory to study transverse spin physics in the hard QCD regime. Such a collider has been proposed by the PAX Collaboration~\cite{PAX-TP} for the new Facility for Antiproton and Ion Research (FAIR) at GSI in Darmstadt, Germany, aiming at luminosities of 10$^{31}$~cm$^{-2}$s$^{-1}$. An integral part of the proposed machine is a dedicated, large--acceptance Antiproton Polarizer Ring (APR).

For more than two decades, physicists have tried to produce beams of polarized antiprotons~\cite{krisch}, generally without success. Conventional methods like Atomic Beam Sources (ABS), appropriate for the production of polarized protons and heavy ions cannot be applied, since antiprotons annihilate with matter. So far the only polarized antiproton beam has been produced from the decay in flight of $\bar{\Lambda}$ hyperons at Fermilab. At polarizations $P>0.35$, the achieved intensities never exceeded $1.5 \cdot 10^5$~s$^{-1}$~\cite{grosnick}. Scattering of antiprotons off a liquid hydrogen target could yield polarizations of $P\approx 0.2$, with beam intensities of up to $2 \cdot 10^3$~s$^{-1}$~\cite{spinka}. Unfortunately, both approaches do not allow efficient accumulation in a storage ring, which would be needed to enhance the luminosity. Spin splitting, using the Stern--Gerlach separation of magnetic substates in a stored antiproton beam was proposed in 1985~\cite{niinikoski}. Although the theoretical understanding has much improved since then~\cite{cameron}, spin splitting using a stored beam has yet to be observed experimentally. In contrast to that, a proof of the spin--filtering principle has been produced by the FILTEX experiment at the TSR--ring in Heidelberg with a proton beam~\cite{TSR}.

The experimental basis for predicting the polarization buildup in a stored antiproton beam by spin--filtering is practically non--existent. The AD--ring at CERN is a unique facility at which stored antiprotons in the appropriate energy range are available and whose characteristics meet the requirements for the first ever antiproton polarization buildup studies. Therefore, it is of highest priority in the quest for polarized antiprotons to make use of this opportunity, and to perform spin--filtering experiments using stored antiprotons at the AD--ring of CERN. In preparation for the experiment at the AD, a number of dedicated spin--filtering experiments will be carried out with protons at the Cooler Synchrotron COSY at J\"ulich, Germany in order to commission the needed equipment, and to gain additional understanding of the accelerator physics aspects of the project.

\cleardoublepage
\section{Physics Case \label{physicscase}}
\pagestyle{myheadings} \markboth{Spin--Dependence
of the $\bar{p}p$ Interaction at the AD}{Physics Case}
\subsection{$N\bar{N}$ Double-Spin Observables from Spin Filtering}
The two double--spin observables, which can be measured by the
spin--filtering technique, are the spin--dependent cross sections
$\sigma_1$ and $\sigma_2$ in the parametrization of the total
hadronic cross section $\sigma_{\mathrm{tot}}$ \cite{bilenky}, written
as
\begin{eqnarray}
\sigma_{\mathrm{tot}}=\sigma_0 + \sigma_1 (\vec{P} \cdot \vec{Q}) +
\sigma_2 (\vec{P} \cdot \hat{k})(\vec{Q}\cdot \hat{k})\;,
\label{eq:sigma-spin-dep}
\end{eqnarray}
where $\sigma_0$ denotes the total spin--independent hadronic cross
section, $\sigma_1$ the total spin--dependent cross section for
transverse orientation of beam polarization $P$ and target
polarization $Q$, $\sigma_2$ denotes the total spin--dependent cross
section for longitudinal orientation of beam and target
polarizations. (Here we use the nomenclature introduced by Bystricky,
Lehar, and Winternitz \cite{bystricky}, where $\hat{k}=\vec{k}/
|\vec{k}|$ is the unit vector along the collision axis.) Such
observables would improve substantially the modern phenomenology of
proton--antiproton interactions based on the experimental data
gathered at LEAR (for a review and references, see \cite{klempt}).

\begin{figure}[hbt]
 \begin{center}
\includegraphics[width=0.65\linewidth]{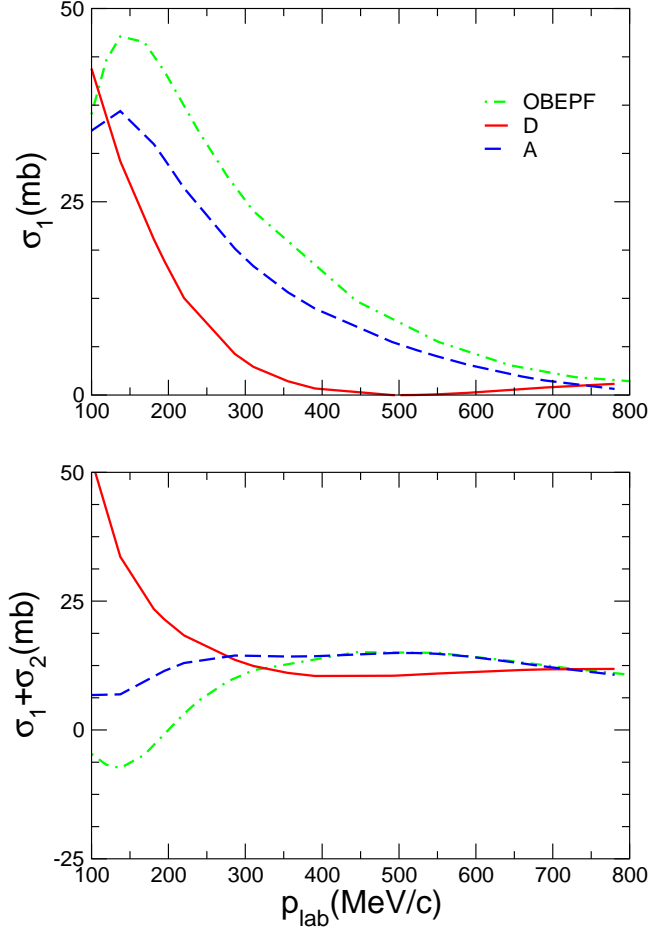}
  \parbox{14cm}{\caption{\label{pbarpmodels}\small The spin--dependent
cross sections $\sigma_1$ and $\sigma_1+\sigma_2$, cf. Eq. (1), as
predicted by the $N\bar N$ models D \cite{Mull}, A \cite{Hippchen} and
OBEPF \cite{Haidenbauer} of the J\"ulich group.
}}
\end{center}
\end{figure}

The suggested spin--filtering experiment at the AD of CERN constitutes
a unique opportunity to measure for the first time these observables
in the 50--450 MeV energy range. The measurements of $\sigma_1$ and
$\sigma_2$ will be carried out in the transmission mode. The
separation of the elastic scattering and annihilation contributions to
$\sigma_1$ and $\sigma_2$ requires the integration of the
double--polarized elastic cross section over the full angular
range. Although such measurements do not seem feasible with the
anticipated luminosity using the former HERMES internal polarized target
installed at the AD, the obtained results on $\sigma_1$ and $\sigma_2$
for the total cross section would serve as an important constraint for
a new generation of baryon--antibaryon interaction models, which will
find broad application to the interpretation of the experimental data
in heavy quark physics. Regarding the main goal of the proposed
experiment -- the antiproton polarization buildup -- the expectations
from the first generation models for double--spin dependence of
$p\bar{p}$ interaction are encouraging, see Fig.~\ref{pbarpmodels}.
With filtering for two lifetimes of the beam, they suggest that in a dedicated large--acceptance storage ring, antiproton beam polarizations of about 15~\% seem within reach (see Fig.~\ref{fig:predictedPol}).

\subsection{$N\bar{N}$ Interaction from LEAR to $J/\Psi$ and $B$-decays}
The evidence for threshold enhancements in $B-$ and $J/\Psi$--decays containing the baryon--antibaryon pairs -- $p\bar{p}$, $p\bar{\Lambda}, \Lambda\bar{p}$, etc. -- was found recently at the modern generation electron--positron colliders BES~\cite{BES1,BES2004,BES2005,BES2005-2}, BELLE~\cite{BELLE1,BELLE2,BELLE3,BELLE4}, and BaBar~\cite{BaBarDecay,BaBarFF}.  These findings added to the urgency of understanding low and intermediate energy $p\bar{p}$ interactions, which appear to be more complex than suggested by the previous analyses \cite{Hippchen,Mull1,Mull,Nijmegen,Paris} of the experimental data from LEAR. The direct measurements of $\sigma_1$ and $\sigma_2$ would facilitate the understanding of the role of antibaryon--baryon final state interactions, which are crucial for the re--interpretation of the B--decay dynamics in terms of the Standard Model mechanisms (see \cite{Rosner,BaBarDecay,Rosner2005} and references therein). Especially strong theoretical activity (\cite{Rosner,He,Bugg,Kerbikov,Sibirtsev,Ding,Datta,Yan,Chang,Chang,Loiseau,Loiseau2,Ding2} and references therein) has been triggered by the BES finding~\cite{BES1} of the pronounced threshold enhancement in the reaction $J/\Psi \to p\bar{p}\gamma$, including the revival of the baryonium states \cite{Bogdanova,Shapiro} in the $p\bar{p}$ system~\cite{Datta,Yan,Chang,Chang,Loiseau,Loiseau2}. Equally important is the recent confirmation by the BaBar collaboration \cite{BaBarFF} of the near--threshold structure in the timelike form factor of the proton, observed earlier at LEAR \cite{LEARFF}, see the theoretical discussion in \cite{HaidenbauerFF,DmitrievFF}. In conjunction with the BES enhancement, the LEAR--BaBar data suggest a non--trivial energy dependence in both the spin--singlet and spin--triplet $p\bar{p}$ interactions, hence our special interest in $\sigma_1$ and $\sigma_2$. Still further interest in the subject was triggered by the recent high statistics data on the $J/\Psi \to p\bar{p}\omega$ decay from the BES collaboration  \cite{BESomega}. The interpretation of these data in terms of the $p\bar{p}$ final state interactions and references to other approaches are found in \cite{HaidenbauerOmega}.

\subsection{Applications of Polarized Antiprotons to QCD Spin
Studies}
The QCD physics potential of experiments with high energy polarized antiprotons is enormous, yet hitherto high luminosity experiments with polarized antiprotons have been impossible. The situation could change dramatically with the demonstration of spin filtering and storing of polarized antiprotons, and the implementation of a double--polarized high--luminosity antiproton--proton collider.  The list of fundamental physics issues for such a collider includes the determination of transversity, the quark transverse polarization inside a transversely polarized proton, the last leading twist missing piece of the QCD description of the partonic structure of the nucleon. It can be directly measured only via double--polarized antiproton--proton Drell--Yan production.  Without measurements of the transversity, the spin tomography of the proton would be ever incomplete. Other items of great importance for the QCD description of the proton include the phase of the timelike form factors of the proton and hard antiproton--proton scattering.  Such an ambitious physics program has been formulated by the PAX collaboration (Polarized Antiproton eXperiment) and a Technical Proposal \cite{PAX-TP} has recently been submitted to the FAIR project.  The uniqueness and the strong scientific merits of the PAX proposal have been well received \cite{PAX-web}, and there is an urgency to convincingly demonstrate experimentally that a high degree of antiproton polarization could be reached with a dedicated APR~\cite{APR}.

\cleardoublepage
\section{Polarizing antiprotons in the AD and proposed measurements\label{sec:polarizing}}
\pagestyle{myheadings} \markboth{Spin--Dependence
of the $\bar{p}p$ Interaction at the AD}{Polarizing antiprotons in the AD and proposed measurements}
\subsection{Methods to polarize antiprotons \label{sec:method}}
When some 25 years ago, intense antiproton beams became a new tool in nuclear and particle physics, there was an immediate demand to polarize these antiprotons. At the time, there was no shortage of rough ideas on how this might be accomplished~\cite{krisch}, but, up to now, the only polarized antiprotons available for use were in a secondary beam facility, which made use of the decay of hyperons~\cite{grosnick}, and which was operating at Fermilab in the 1990’s.

Only a few of the early ideas to polarize antiprotons are still considered today~\cite{daresbury}, and they all aim at polarizing an ensemble of antiprotons that is orbiting in a storage ring. Spin--$\frac{1}{2}$ particles have two possible spin projections. Polarizing a beam of such particles could be accomplished by identifying and selectively discarding the particles in one of the two states. The separation of the spin states could be accomplished via the interaction of the magnetic moments with external fields (known as ‘Stern-Gerlach effect’), or via the spin dependence of a nuclear reaction (known as ‘spin filtering’). While there is doubt that the former scheme is feasible, the second method has actually been subjected to a successful experimental test~\cite{TSR}.

A spin--$\frac{1}{2}$ beam could also be polarized if particles in one spin state would be moved into the other state (by ‘spin flipping’). The advantage over the spin filter method is that the precious stored beam is conserved by this process. Some time ago it was  suggested that such a spin flip may be induced by an interaction of the stored beam with polarized electrons. Recently, the Mainz group~\cite{arenhoevel} calculated a very large spin--flip probability in electron--proton scattering at low energy (around 1 eV center--of--mass energy). The predicted cross sections were so large that it would have become possible to polarize stored antiprotons by spin flip, using a co--moving beam of polarized positrons~\cite{walcher} even at the low intensities at which such a beam is currently available. However a competing calculation~\cite{milstein08} disagreed with the estimate of the Mainz group by 16 orders of magnitude and came to the conclusion that spin flip in hadron--lepton scattering is not a viable mechanism to polarize a stored beam. The discrepancy between the two theoretical estimates was resolved by a very recent, dedicated experiment at COSY in which the depolarization of a stored, polarized proton beam scattering from unpolarized electrons, was measured~\cite{oellers}. This experiment establishes an upper limit for the spin--flip cross section that clearly rules out the estimate by the Mainz group~\cite{arenhoevel,walcher}, and for the time being the prospect of using spin flip to polarize a stored beam, at least until a better idea provides new hope. Since the completion of the experiment, the calculations of the Mainz group have  been withdrawn~\cite{arenhoevel2,walcher2}.

Thus, at this time, spin filtering is the only known method that stands a reasonable chance of succeeding in the production of a stored beam of polarized antiprotons.
\subsection{Spin filtering and measurement of spin dependence of the $\bar{p}p$ interaction   \label{technique}}
That a stored beam can be polarized by spin filtering has been demonstrated in the FILTEX experiment at TSR, Heidelberg~\cite{TSR}. In this experiment, a 23--MeV, stored proton beam was passing through an internal transversely polarized hydrogen gas target with  a thickness of $6 \times 10^{13}$ atoms/cm$^2$. The polarization buildup of the beam as a function of filter time $t$ can be expressed as~\cite{TSR}
\begin{eqnarray}
P(t)=\tanh(t/\tau_1)\;,
\end{eqnarray}
provided that the effect of depolarizing resonances can be neglected on a time scale of $\tau_1$. Here, $\tau_1$ is the polarizing time constant. The induced beam polarization is parallel to the target polarization. In FILTEX, a polarization buildup rate of 
\begin{eqnarray}
\frac{\mathrm{d}P}{\mathrm{d}t}  \approx \frac{1}{\tau_1} = 0.0124 \pm  0.0006 \; \mathrm{per\; hour}
\end{eqnarray}
has been observed. 

The polarization buildup has been interpreted in terms of the known $pp$ spin--dependent interaction (see, e.g., \cite{milstein,heimbach}). It can be shown that the time constants for transverse ($\bot$) or longitudinal ($\parallel$) filtering are given by, respectively  
\begin{eqnarray}
\tau_1^{\bot}=\frac{1}{\tilde{\sigma}_1 Q d_t f} \;\;\; \mathrm{and} \;\;\;
\tau_1^{||}=\frac{1}{(\tilde{\sigma}_1+\tilde{\sigma}_2) Q d_t f}\;.
\label{eq:tau}
\end{eqnarray}
Here, $Q$ is the target polarization, $d_t$ is the target thickness in atoms/cm$^2$ and $f$ is the revolution frequency of the particles in the ring.  The “filtering cross sections”   $\tilde{\sigma}_1$ and $\tilde{\sigma}_2$̃ are closely related to the spin--dependent total cross sections, $\sigma_1$  and $\sigma_2$, defined in Eq.~(\ref{eq:sigma-spin-dep}). The difference arises because antiprotons that scatter at a sufficiently small angle remain in the ring. This is the case for scattering events with $\theta$ less than the acceptance angle $\theta_\mathrm{acc}$ of the machine downstream of the target.

In order to extract both spin--dependent total cross sections $\sigma_1$ and $\sigma_2$ from the observed time constants [Eq.~(\ref{eq:tau})], a measurement with transverse and longitudinal target (and therefore beam) polarization is required. The latter involves the operation of a Siberian snake in the AD. In addition, the acceptance angle $\theta_\mathrm{acc}$ has to be known (typically 10 -- 20~mrad, see Sec.~\ref{sec:acceptance-lifetime}), because the spin--dependent cross sections $\tilde{\sigma}_1$ and $\tilde{\sigma}_2$ depend on $\theta_{\rm acc}$. Therefore, Coulomb--nuclear interference at extreme forward angles $\theta < \theta_\mathrm{acc}$ must be taken into account. For protons, where the interaction is purely elastic, this is accomplished using the existing $NN$ scattering databases SAID~\cite{said} and Nijmegen~\cite{nijmegen}. For antiprotons such a data base does not exist yet, and since both elastic scattering and all annihilation channels contribute, the corrected cross sections need to be calculated within a specific model.

Recently, a calculation was published of the spin--dependent cross sections in $\bar{p}p$, $\bar{p}n$, and $\bar{p}d$ at intermediate energy, based on $\bar{N}N$ interaction models developed by the J\"ulich group~\cite{uzikov}. While the unpolarized total cross sections reproduce the data quite well, the predicted spin--dependence depends strongly on the interaction model chosen, emphasizing the need for empirical information. It was also found that shorter buildup time constants might be achieved using a polarized deuterium target. The polarized target to be installed at the AD would be capable of providing polarized deuterium, so this option will be seriously considered. 

In addition to a test of the feasibility of polarizing antiprotons by filtering, the proposed AD experiment will provide a measurement of the spin--dependent cross sections $\sigma_1$ and $\sigma_2$ as a function of energy from 50 to 450~MeV. First information will be obtained on the spin--dependence of the antiproton--proton and the antiproton--deuteron interaction. These data will represent a crucial test of current and emerging theoretical models.

\subsection{AD machine properties}
\subsubsection{Acceptance and beam lifetime expectation\label{sec:acceptance-lifetime}}
The stored beam intensity decreases with time because of reactions and scattering outside the machine acceptance. Most of these events occur in the internal target. The beam lifetime depends thus on the target thickness. For a given production rate of polarized atoms, the target thickness depends on the diameter of the storage cell. The opening in the storage cell, in turn, affects the machine acceptance, and one is faced with an optimization problem. Possible optimization criteria are the polarization buildup rate, the beam remaining after filtering, or the statistical accuracy of a polarization measurement.

\begin{figure}[hbt]
 \begin{center}
\includegraphics[width=0.45\linewidth]{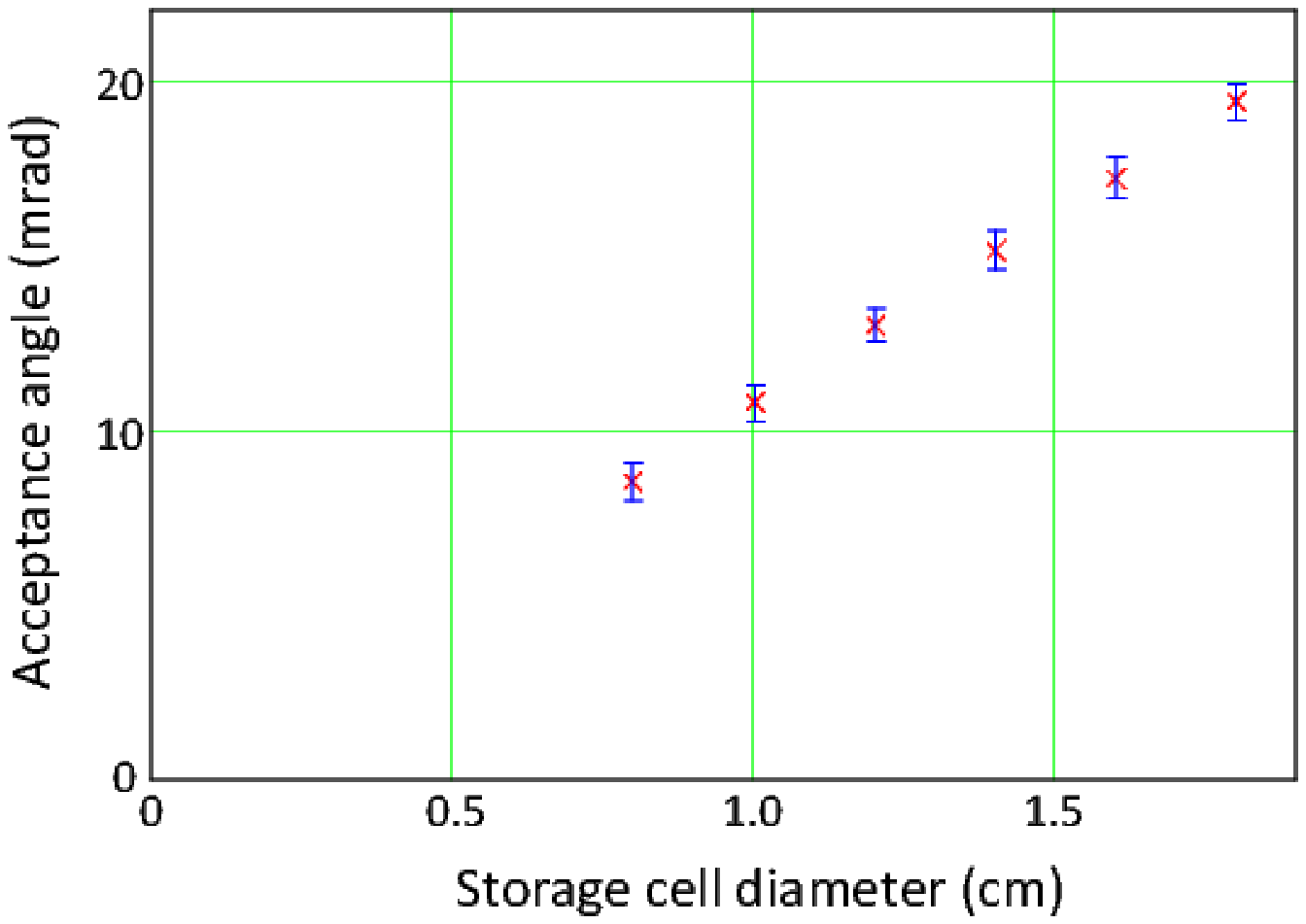}
\includegraphics[width=0.45\linewidth]{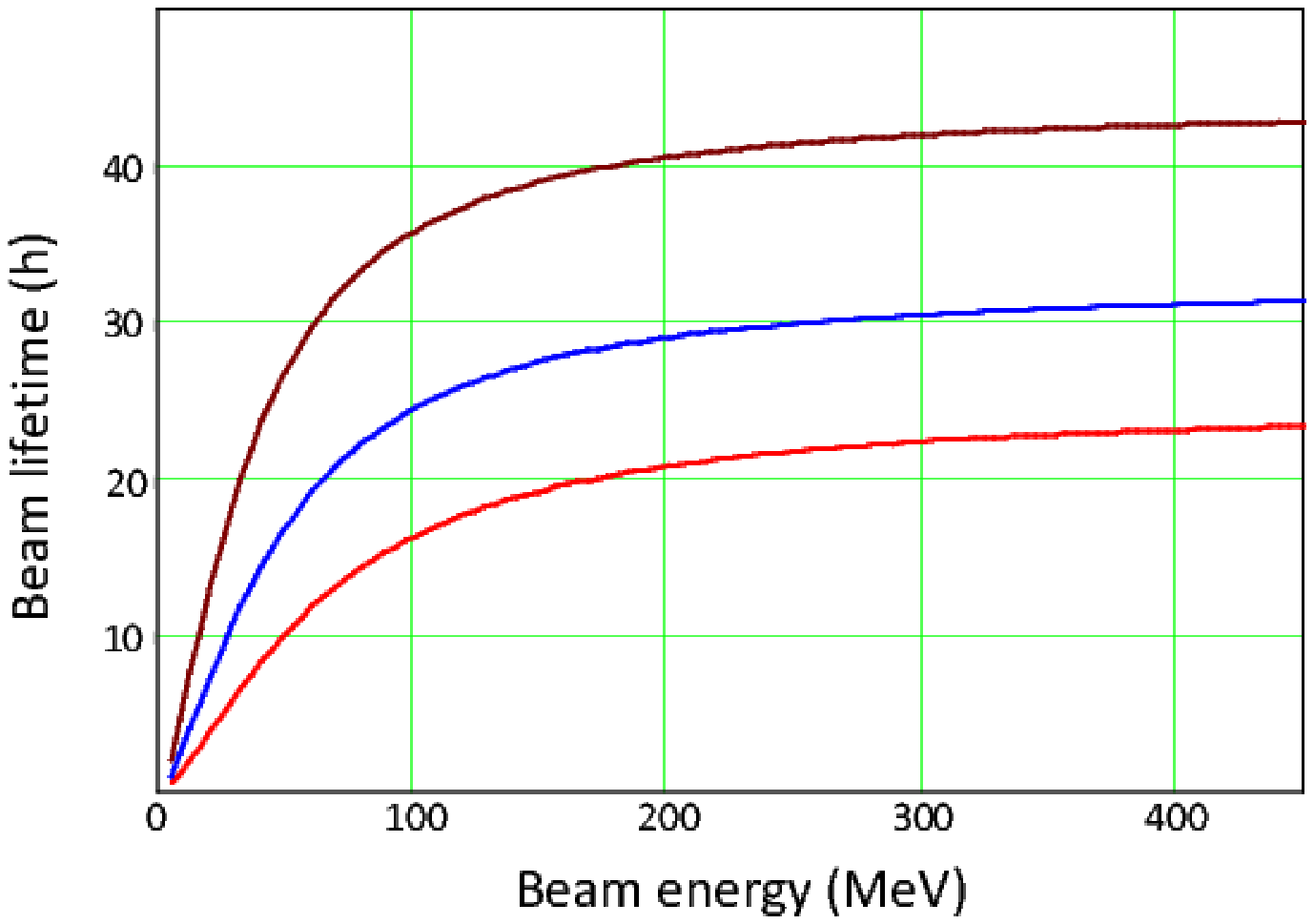}
  \parbox{14cm}{\caption{\label{fig:thetavsd}\small 
Left panel: Acceptance angle $\theta_\mathrm{acc}$ versus the diameter of the storage cell . Right panel: Estimated beam lifetime as a function of the beam energy for the storage cell diameters 0.8~cm ({\color{red}red}), 1.0~cm, ({\color{blue}blue}), and 1.2~cm ({\color{Brown} brown}).
}}
\end{center}
\end{figure}
As an example, the left panel of Fig.~\ref{fig:thetavsd}  shows the dependence of the acceptance angle in the AD as a function of the diameter of a storage cell at the PAX target. This relation uses a machine acceptance of $A_x = 200$~ $\mu$m and $A_y = 180$~$\mu$m and a $2\sigma$ beam emittance of 1~$\mu$m in both planes~\cite{tranquille}. The right panel shows predicted beam lifetimes for three different cell diameters. It is assumed that the beam losses occur entirely in the target, taking into account Rutherford scattering and the $\bar{p}p$ total cross section. 

\subsubsection{Beam intensity\label{sec:beam-intensity}}
At present, the AD provides about $3 \times 10^7$ stored antiprotons. We anticipate that by the development of stacking techniques one may be able to increase this number by about a factor of five~\cite{cern-talk}. With  about $10^8$ stored antiprotons, a luminosity of $L = N_{\bar{p}}  \cdot f \cdot d_t$ $= 10^8 \cdot  10^6$~s$^{-1} \cdot  10^{14}$ atoms/cm$^2$ $= 10^{28}$~cm$^{-2}$s$^{-1}$ will be achievable.
 
Keeping in mind that after spin filtering for a few beam lifetimes  a substantial remaining beam intensity is needed in order to measure the beam polarization, it is fairly important to dedicate some development effort towards increasing the number of antiprotons stored in the AD. Obviously, other experiments at the AD would also benefit from such a development.
\subsubsection{Polarization lifetime expectation}
The spin tune is the number of rotations of the magnetic moment of a particle during one turn around the ring. Depolarizing resonances arise when the horizontal and vertical tune, the orbit frequency and the synchrotron frequency, or combinations thereof, are related in a simple way to the spin tune. As an example, an ‘intrinsic’ resonance occurs when the (e.g.) horizontal tune and the spin tune are an integer multiple of each other. More complicated resonance conditions are possible and lead to (usually weaker) resonances. Near a resonance, the spin motions of individual particles start to differ, and the beam depolarizes.

Experimental studies of the polarization lifetime in a storage ring have been carried out at the Indiana Cooler~\cite{meyer,przewoski}. It was found that the polarization lifetime increases rapidly with distance from the resonance, and quickly becomes so long as to be difficult to measure. Thus, for the Cooler experiments it was sufficient to avoid the immediate proximity to any low-order resonance.

For experiments involving long filtering times, the conditions on the preservation of polarization are more stringent and higher--order resonances could become an issue. We anticipate that in preparation for the experiments at the AD a careful study of depolarizing resonances will be needed, in order to guarantee that the polarization lifetime is much longer than the buildup time.

\subsection{Anticipated polarization buildup in the AD \label{sec:buildup}}
In this section we present an estimate of the anticipated polarization buildup by filtering. For the spin--dependent $\bar{p}p$ interaction we are using two different models (models “A” and “D” of ref. \cite{uzikov}). 
\begin{figure}[!]
 \begin{center}
\includegraphics[width=0.49\linewidth]{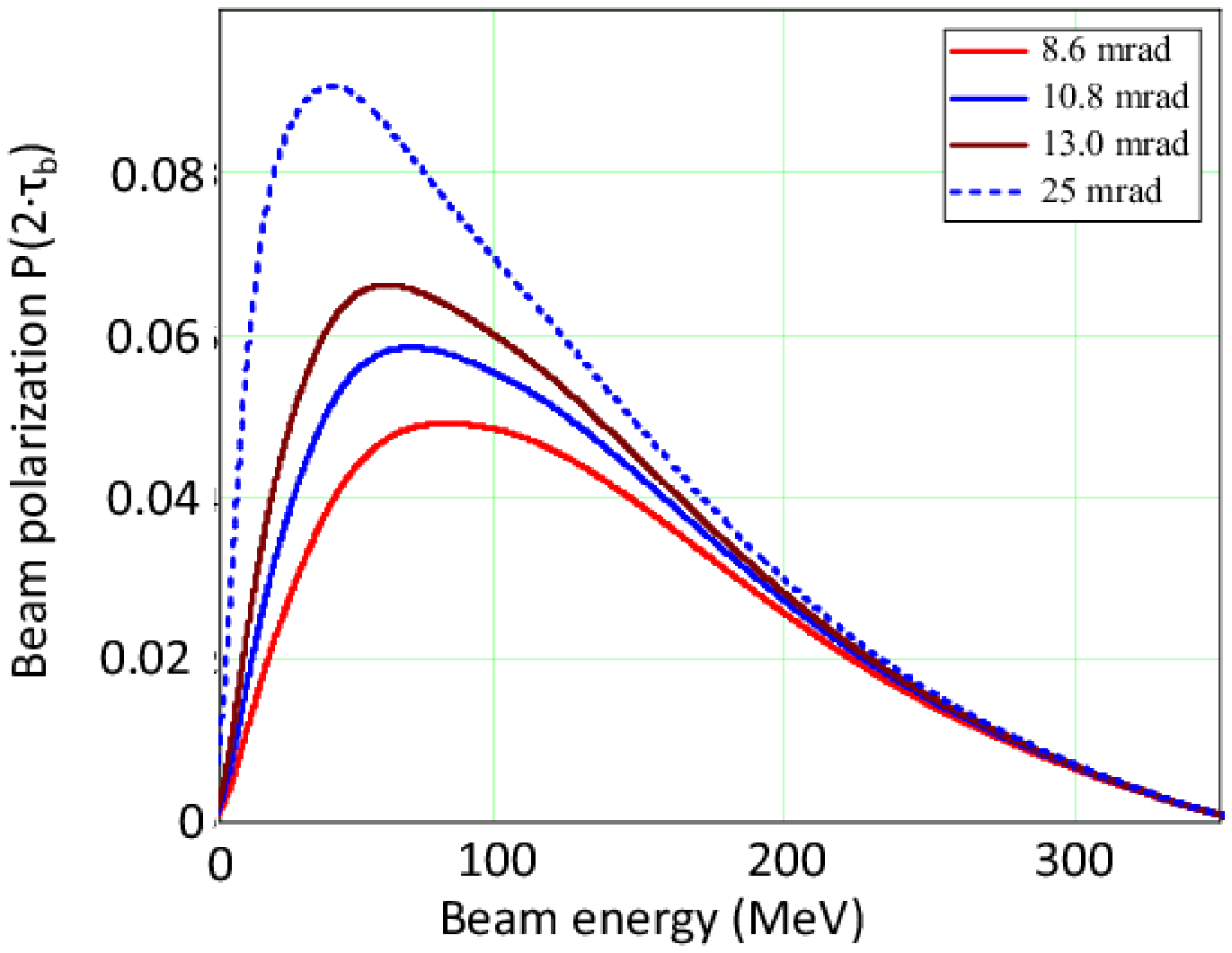}
\includegraphics[width=0.49\linewidth]{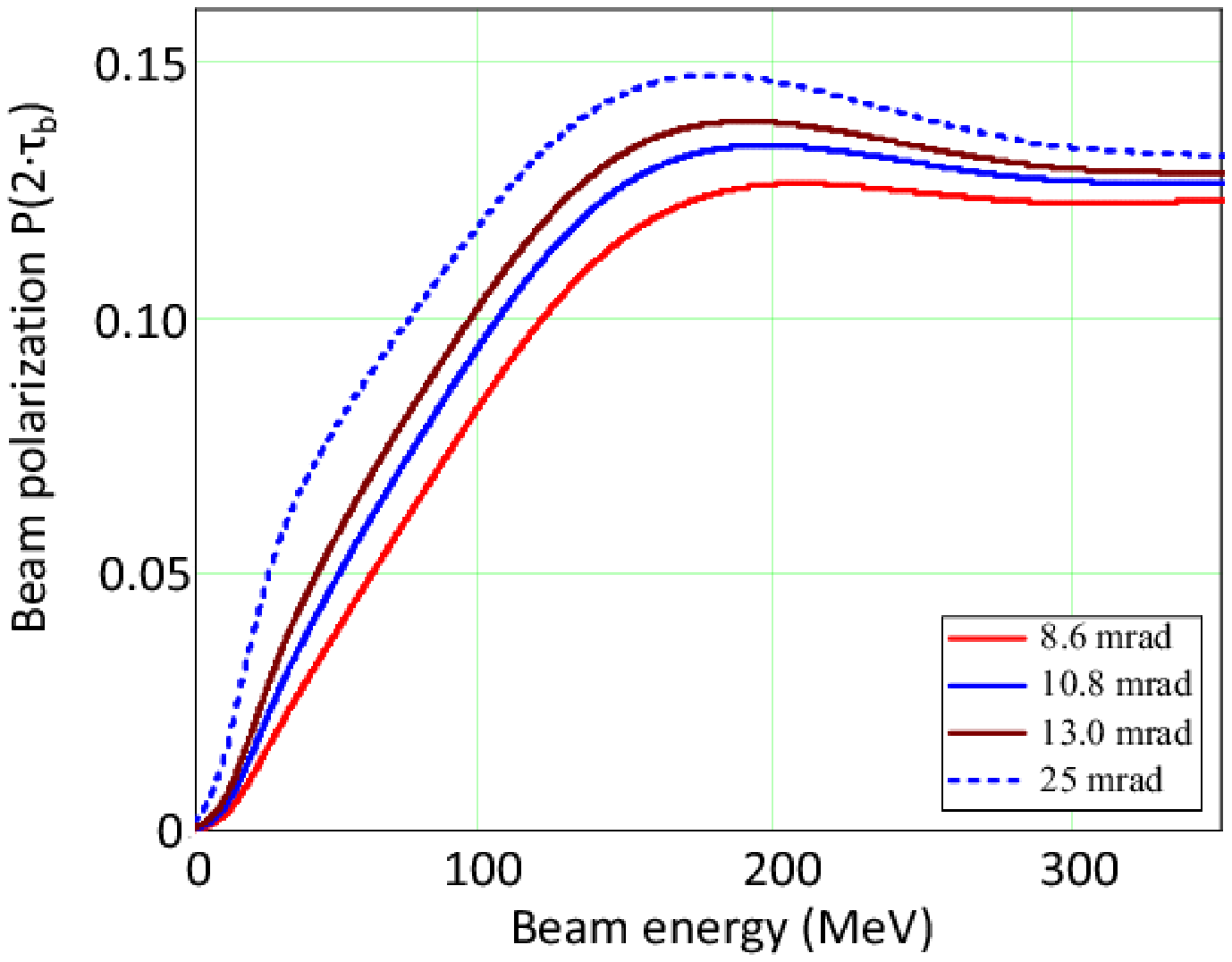}
\includegraphics[width=0.49\linewidth]{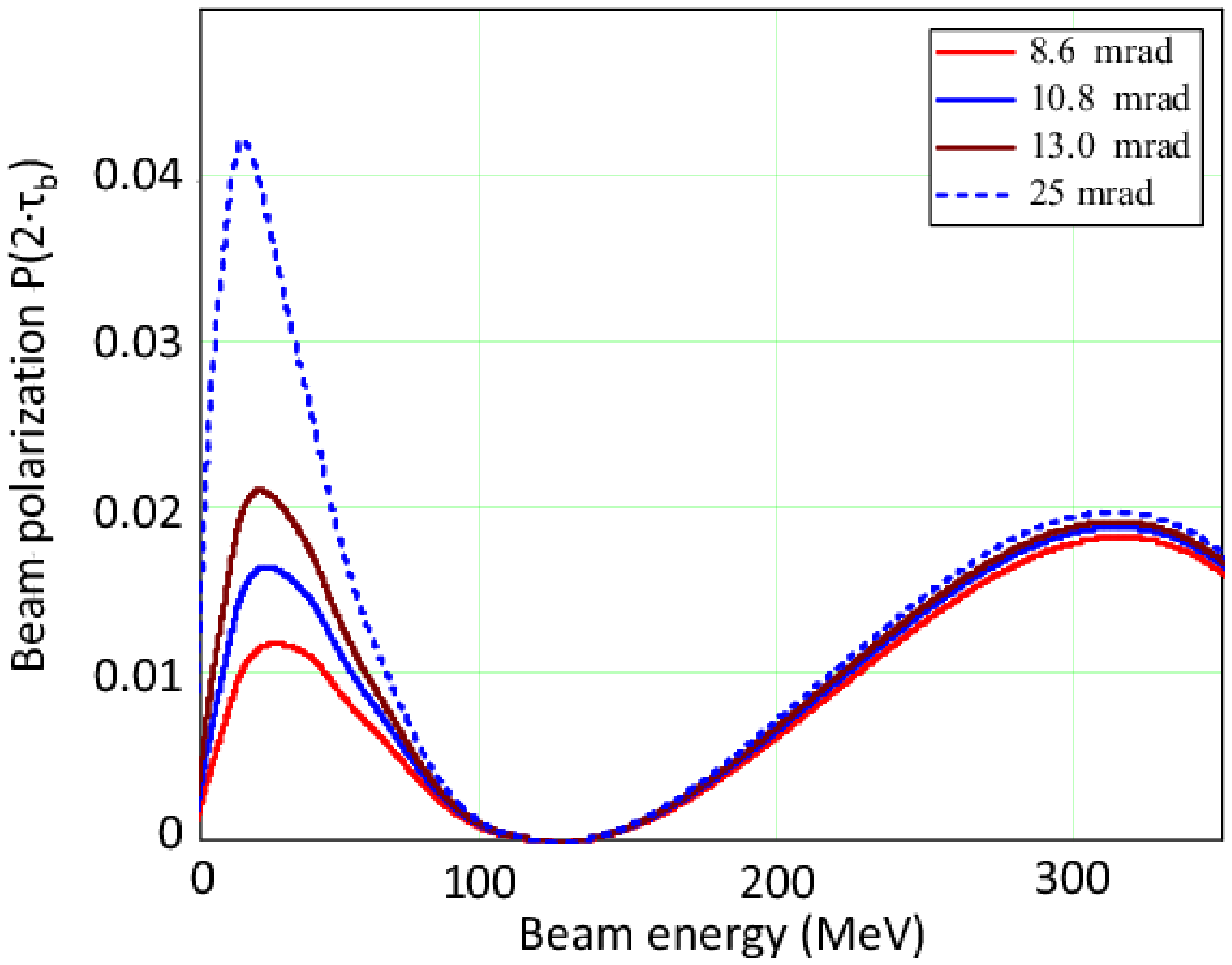}
\includegraphics[width=0.49\linewidth]{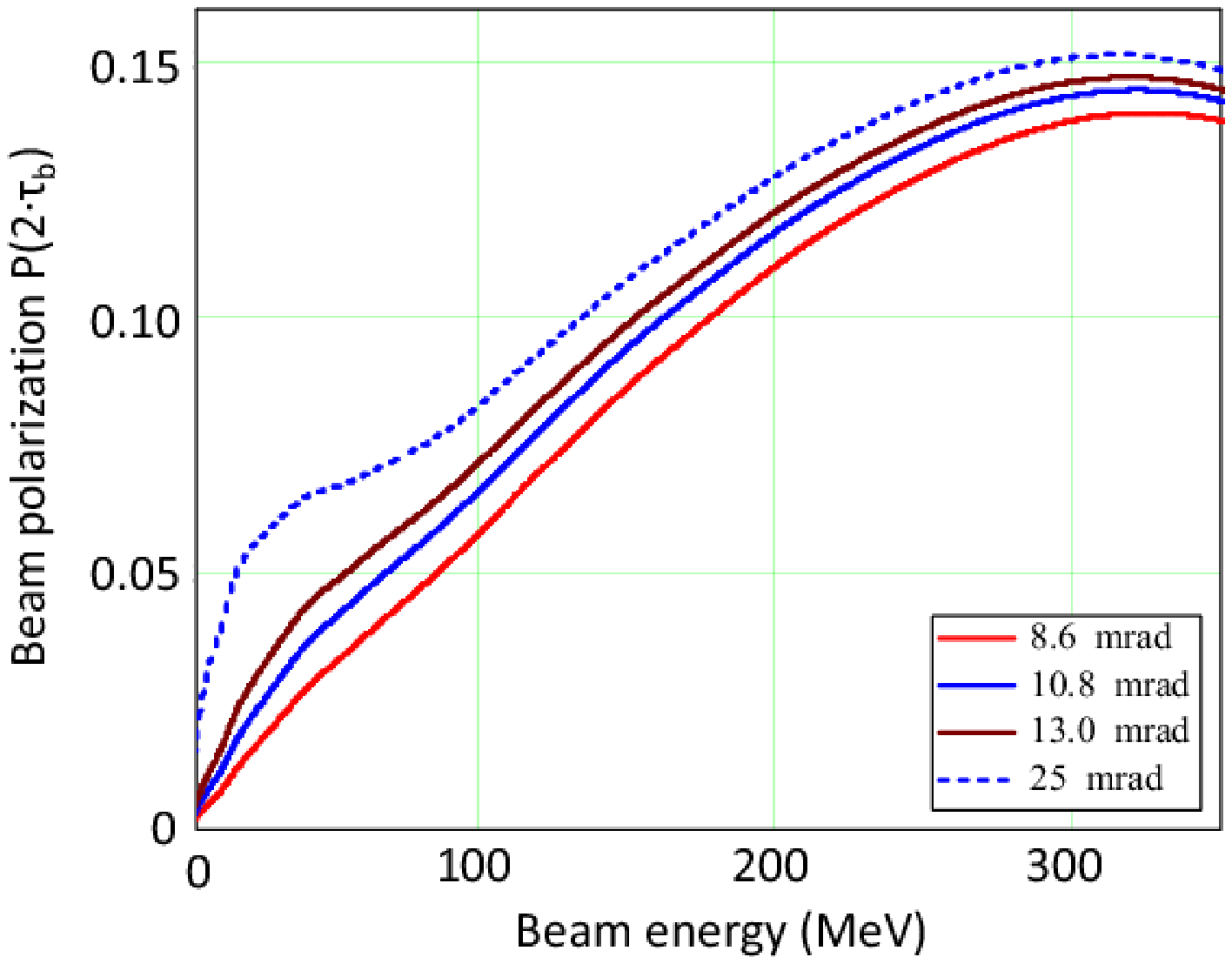}
  \parbox{14cm}{\caption{\label{fig:predictedPol}\small 
Estimated beam polarization after two beam lifetimes of filtering (corresponding to a factor of about 7 decrease in the number of stored antiprotons) as function of the beam energy. The panels on the left and on the right are for transverse and longitudinal polarization, respectively.  The panels in the top and bottom rows are for models “A” and “D”~\cite{uzikov}, respectively. The solid lines correspond to different cell diameters (see caption, Fig.~\ref{fig:thetavsd}).
}}
\end{center}
\end{figure}
The solid lines are calculated for different cell diameters, and correspond to the curves in the right panel of Fig.~\ref{fig:predictedPol}; the dashed line is for the theoretical acceptance angle limit of 25 mrad. 

\cleardoublepage
\section{Polarimetry of beam and target \label{sec:polarimetry}}
\pagestyle{myheadings} \markboth{Spin--Dependence
of the $\bar{p}p$ Interaction at the AD}{Polarimetry of beam and target}
The primary quantity measured in this experiment is the polarization of the stored beam. 

\begin{figure}[b]
\begin{center}  
\includegraphics[width=0.68\linewidth]{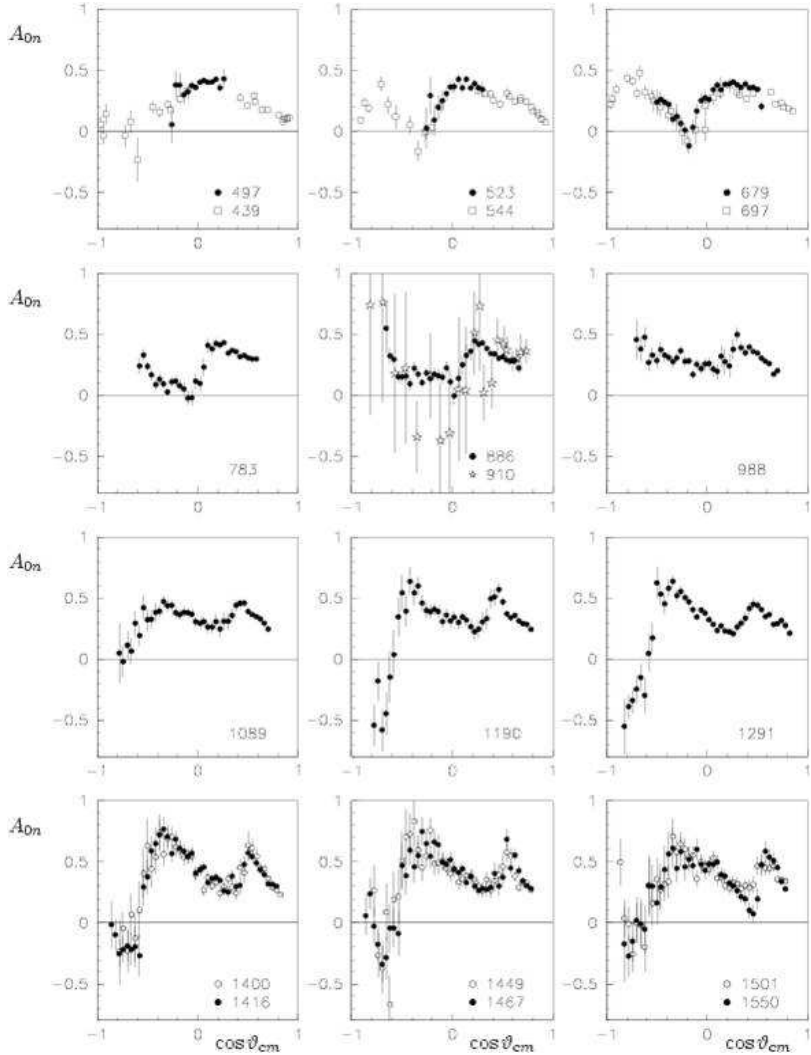}
 \parbox{14cm}{\caption{\label{fig:pbarpAy}\small Analyzing powers in $\bar{p}\vec{p}$ scattering~\cite{bertini,kunne,perrot-kunne}.}} 
\end{center}
\end{figure}

Vertical beam polarization is measured by observing the left--right $\bar{p}p$ scattering asymmetry. This asymmetry is given by the product of the beam polarization and the analyzing power. The $\vec{\bar{p}}p$ analyzing power, of course, has never been measured but is related to the $\bar{p}\vec{p}$ analyzing power by CPT symmetry. In our energy range, angular distributions of the $\bar{p}\vec{p}$ analyzing power have been measured at LEAR by experiment PS172~\cite{kunne} at fifteen momenta, ranging from 497 to 1550 MeV/c, and by experiment PS198 at 439, 544, and 697~MeV/c~\cite{bertini,perrot-kunne}. Most of the data are shown in Fig.~\ref{fig:pbarpAy}. 
It is important to realize that our experiment generates its own $\bar{p}\vec{p}$ analyzing powers, since, at least during the early part of the filtering process, the (still) unpolarized antiproton beam is interacting with a hydrogen target of known, large polarization. 

In order to avoid contributions from other spin observables, it is assumed that during the polarization measurement the target is unpolarized. This is easily achieved by averaging data taken with spin up and down, or by injection of unpolarized H$_2$ into the cell (see Sec.~\ref{sec:ugfs}).

Longitudinal beam polarization could only be directly measured when the spin--corre\-lation coefficients were known. Thus, in this case, before measuring the beam polarization, the spin alignment axis at the target must be turned from longitudinal to vertical by adiabatically turning off the Siberian snake. It is interesting to note that with the polarized internal target, a determination of the longitudinal spin correlation coefficient $A_{zz}$ in elastic $\vec{\bar{p}}\vec{p}$ scattering becomes possible. Once this parameter is established, the longitudinal beam polarization could be determined directly.

The target polarization will either be measured using known $\bar{p}p$ analyzing powers or by the Breit--Rabi polarimeter that analyzes a fraction of the atomic beam that is generated in the target source (see Sec.~\ref{sec:tga-brp}). The target density can be either obtained from the observed deceleration of the stored beam when the electron cooling is switched off, as shown in refs.~\cite{stein,zapfe}, or it can be inferred from the measured rates in the polarimeter using the quite well established elastic antiproton--proton differential cross sections, measured at LEAR~\cite{klempt}.

The detector system needed to observe $\bar{p}p$ scattering is quite  similar to the one that has recently been developed for the ANKE spectrometer operated at COSY~\cite{schleichert}. This system uses microstrip detectors in close proximity to the luminous target volume (see ref.~\cite{oellers}, and also Sec.~\ref{sec:detector}).

\cleardoublepage
\section{Experimental setup for the AD--ring \label{sec:setup}}
\pagestyle{myheadings} \markboth{Spin--Dependence of the
$\bar{p}p$ Interaction at the AD}{Experimental setup for the AD--ring}
At present, the AD of CERN is  the only place world wide, where the proposed measurements can be performed. The effort involved is substantial. Although we will perform most of the design and commissioning work outside of CERN many aspects in the design require a close collaboration with the CERN machine group. The new components that need to be installed in the AD are described in the following sections. They shall all be tested and commissioned at the Cooler Synchrotron COSY in J\"ulich. During these tests we plan to perform dedicated spin--filtering experiments with protons.

\subsection{Overview}
\begin{figure}[hbt]
 \begin{center}
\includegraphics[width=0.8\linewidth]{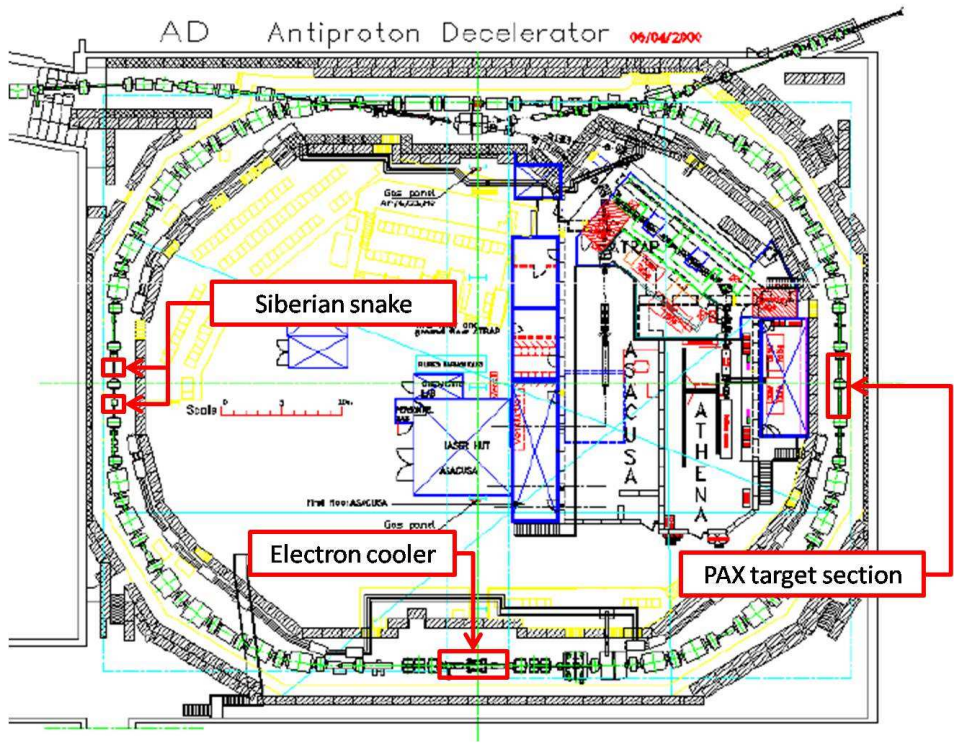}
  \parbox{14cm}{\caption{\label{fig:AD-layout}\small 
Floorplan of the AD showing the foreseen locations of the polarized internal target, the upgraded electron cooler and the Siberian snake. 
}}
\end{center}
\end{figure}

The measurement requires implementing a Polarized Internal storage cell Target (PIT) in the straight section between injection and electron cooling of the AD (see Fig.~\ref{fig:AD-layout}). 
Polarized internal targets represent nowdays a well established technique with high performance and reliability shown in many different experiments with hadronic and leptonic probes. Targets of this kind have been operated successfully at TSR in Heidelberg~\cite{filtex-target}, later on they were also used at HERA/DESY~\cite{hermes-target}, at Indiana University Cyclotron Facility and at MIT--Bates. A new PIT is presently operated at ANKE--COSY \cite{SPIN,grigoriev2}. A recent review can be found in ref.~\cite{steffens-haeberli}. Typical target densities range from a few 10$^{13}$ to $2\times10^{14}$~atoms/cm$^2$~\cite{hermes-target}. The 
\begin{landscape}
\begin{figure}[hbt]
\vspace{-2cm}
 \begin{center}
\includegraphics[width=\linewidth]{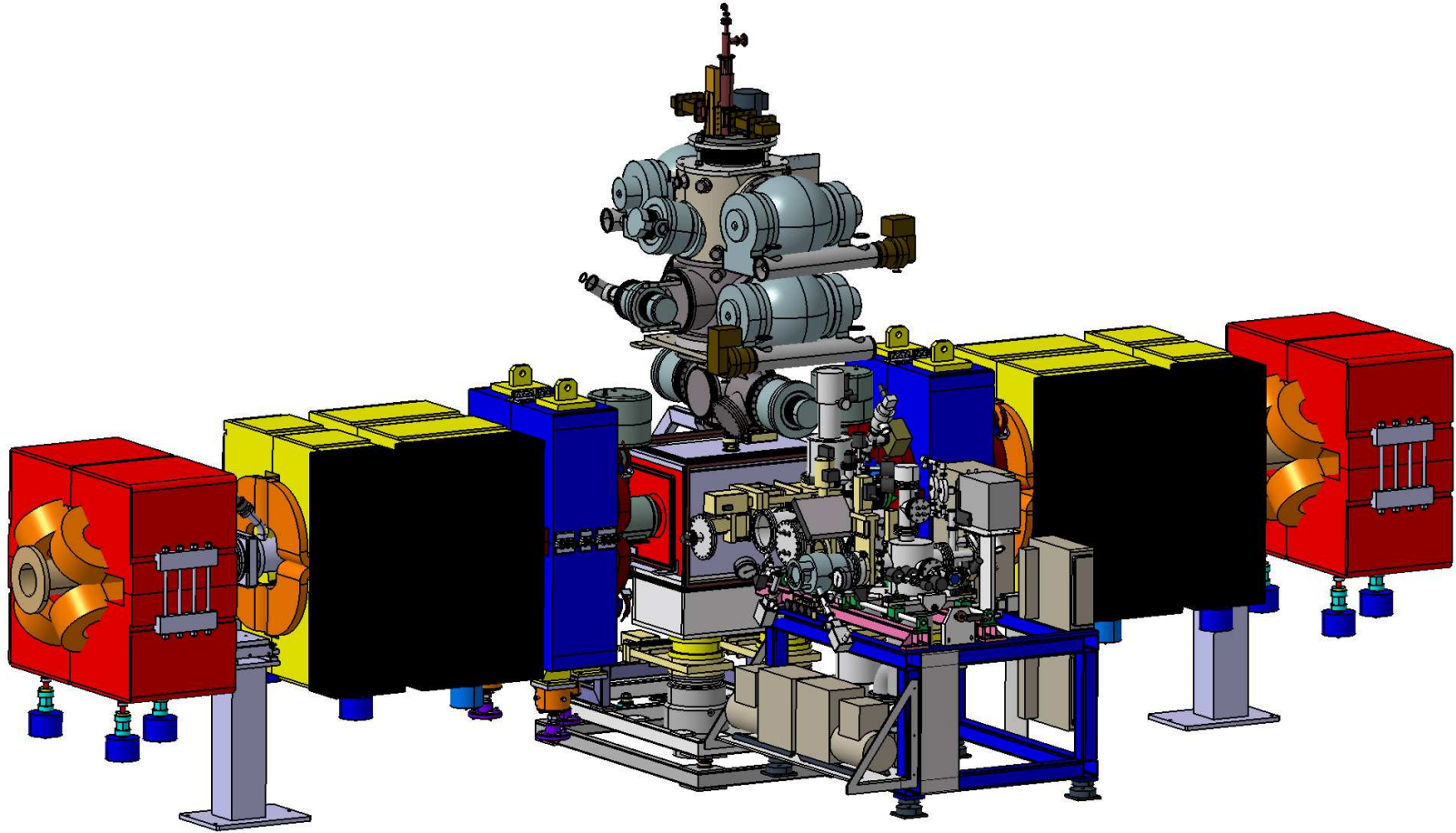}
  \parbox{22cm}{\caption{\label{fig:setup-overview}\small Full installation foreseen at the AD for the straight section between injection and electron cooling. (The beam moves from right to left.) Shown in red are the existing AD quadrupole magnets that define the up-- and downstream boundaries of the low--$\beta$ insertion. The magnets show in yellow are spare COSY straight section (long) and arc (short)  quadrupole magnets. Two additional short quadrupoles have been recuperated from the CELSIUS ring (blue). The atomic beam source is mounted above the target chamber that houses the detector system and the storage cell. Three sets of Helmholtz coils  providing magnetic holding fields along $x$, $y$, and $z$ are mounted on the edges of the target chamber (brown). The Breit--Rabi target polarimeter and the target--gas analyzer are mounted outwards of the ring. Fast shutters are used on the target chamber on all four main ports. The complete section can be sealed off from the rest of the AD by valves (installed between the yellow and red quadrupole magnets).}}
\end{center}
\end{figure}
\end{landscape}
\noindent target density depends strongly on the transverse dimension of the storage cell. In order to provide a high target density, the $\beta$--function at the storage cell should be about $\beta_x=\beta_y=0.3$~m.  In order to minimize the $\beta$--functions at the cell, a special insertion is proposed, which includes additional quadrupoles around the storage cell (see Fig.~\ref{fig:setup-overview}). The low--$\beta$ section should be designed in such a way that the storage cell limits the machine acceptance only marginally. A careful machine study has been carried out in order to maintain the machine performance at injection energy and at low energies for the other AD experiments. 

We will utilize the PIT formerly used at HERA/DESY, which has already become available at the beginning of 2006, to feed the storage cell. The target will be operated in a weak magnetic guide field of a about 10~G. The orientation of the target polarization can be maintained by a set of Helmholtz coils in transverse and longitudinal directions.

\subsection{Low--$\beta$ section\label{sec:low-beta}}
\begin{figure}[b]
 \begin{center}
\includegraphics[width=0.49\linewidth]{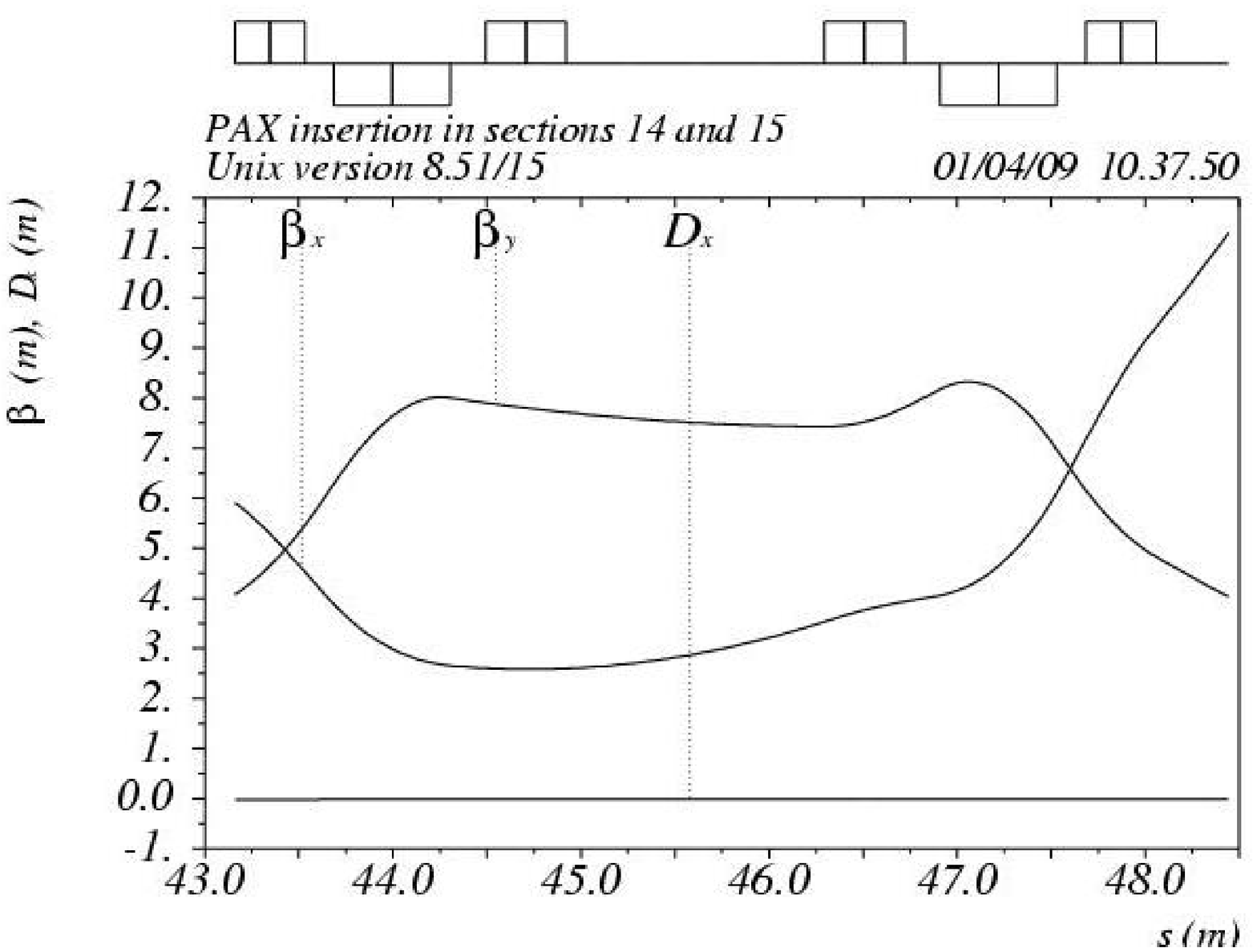}
\includegraphics[width=0.49\linewidth]{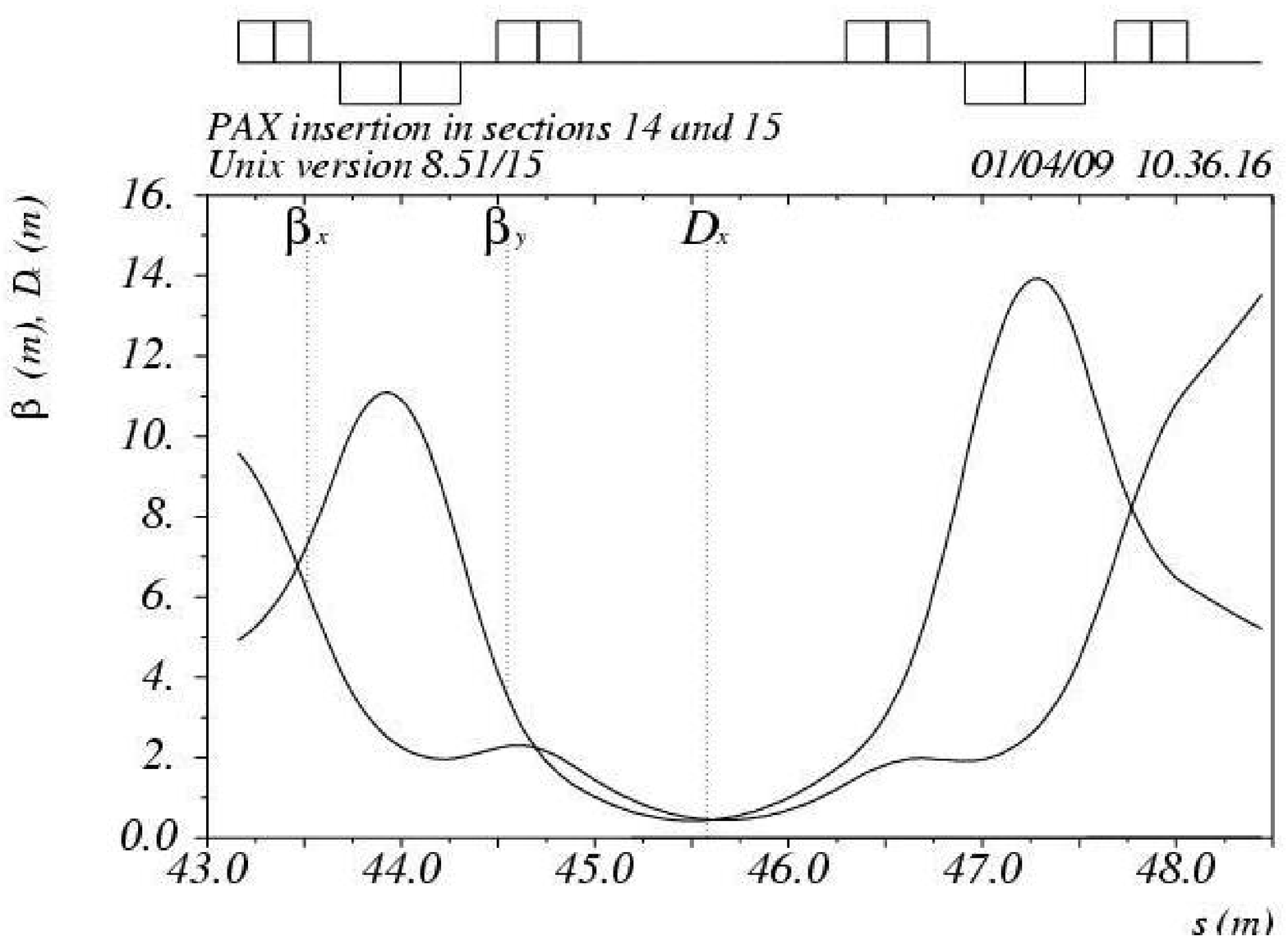}
  \parbox{14cm}{\caption{\label{fig:lattice}\small 
Twiss functions and dispersion in the low--$\beta$ insertion for the PAX installation at the AD. The machine optics at injection energy (3.57~GeV/c) is shown in the left panel. After the beam has been ramped down to the experiment energies in the range of 50--450 MeV, the machine optics is squeezed (right panel) by fully powering on the additional six quadrupole magnets (see Fig.~\ref{fig:setup-overview}) for operation of the polarized target. The center of the storage cell is located at 45.608~m.
}}
\end{center}
\end{figure}

The operation of the polarized target requires transporting the stored beam through the narrow storage cell. Since at injection into the AD at 3.57~GeV/c, the beam is not yet cooled, the apertures in the target region shall not be restricted by the storage cell. For the measurements proposed here, the beam is ramped down to the energies of interest (50--450~MeV). There, the machine optics is \textit{squeezed} by applying stronger focussing using the additional quadrupole magnets shown in Fig.~\ref{fig:setup-overview}, and only after this is accomplished, the storage cell is closed. The two situations are depicted in Fig.~\ref{fig:lattice} in terms of the betatron amplitudes in the low--$\beta$ insertion.

\begin{figure}[t]
 \begin{center}
\includegraphics[angle=-90,width=0.8\linewidth]{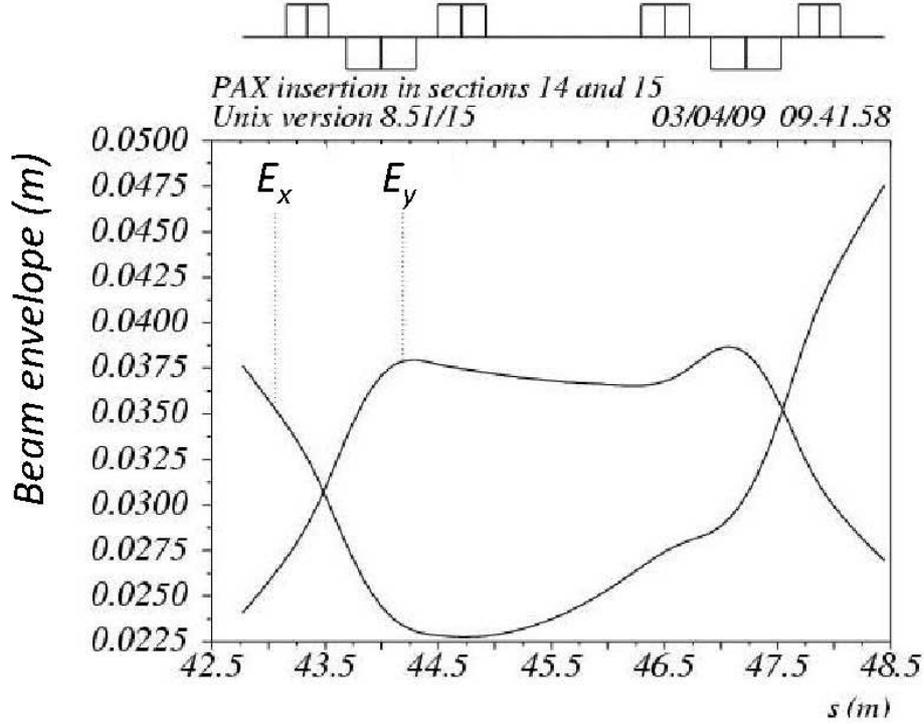}
  \parbox{14cm}{\caption{\label{fig:envelope}\small Beam envelope in the PAX low--$\beta$ insertion when the beam is injected into the AD. The center of the storage cell is located at 45.608~m.
}}
\end{center}
\end{figure}

The operation of the AD implies that the storage cell must be opened at injection until the beam is ramped down to experiment energy and the 'squeezed' optics is set up. In Fig.~\ref{fig:envelope}, the beam envelopes $E(x)$ and $E(y)$ are shown based on the $\beta$--functions from Fig.~\ref{fig:lattice} (left panel) using $E(x,y)=\sqrt{A_{x,y} \cdot \beta_{x,y}}$ and the nominal AD machine acceptances of $A_x=200$~$\mu$m and $A_y=180$~$\mu$m.

Prior to the development of a solution of the low--$\beta$ section based on normal--conducting quadrupole magnets, shown in Fig.~\ref{fig:setup-overview}, the PAX collaboration spent a considerable effort on the development of superconducting (SC) quadrupole  magnets. The main disadvantage of SC magnets is that for their application at the AD, they would need to be ramped as quickly as the normal--conducting magnets in order not to affect adversely the normal AD operation with its fast downramp. Otherwise, the only feasible solution would have been to install the complete low--$\beta$ section for the PAX measurements, and to de--install it afterwards. A brief description of the development work on SC quadrupole magnets can be found in ref.~\cite{sc-quadrupoles}. We are currently building a short SC racetrack coil for a prototype of such a quadrupole magnet.

\subsection{Polarized internal target}
The PAX experiment will make use of the target previously employed at HERMES, an experiment at HERA--DESY~\cite{hermes-target} which stopped operating the target in 2005 and concluded unpolarized data taking in 2007. Meanwhile, the target and the polarimeter have been relocated  to J\"ulich and put back into operation. At COSY, a number of spin--filtering measurements with protons will be carried out, and the experimental setup for AD experiments proposed here will be commissioned.

\begin{figure}[t]
 \begin{center}
\includegraphics[width=0.42\linewidth]{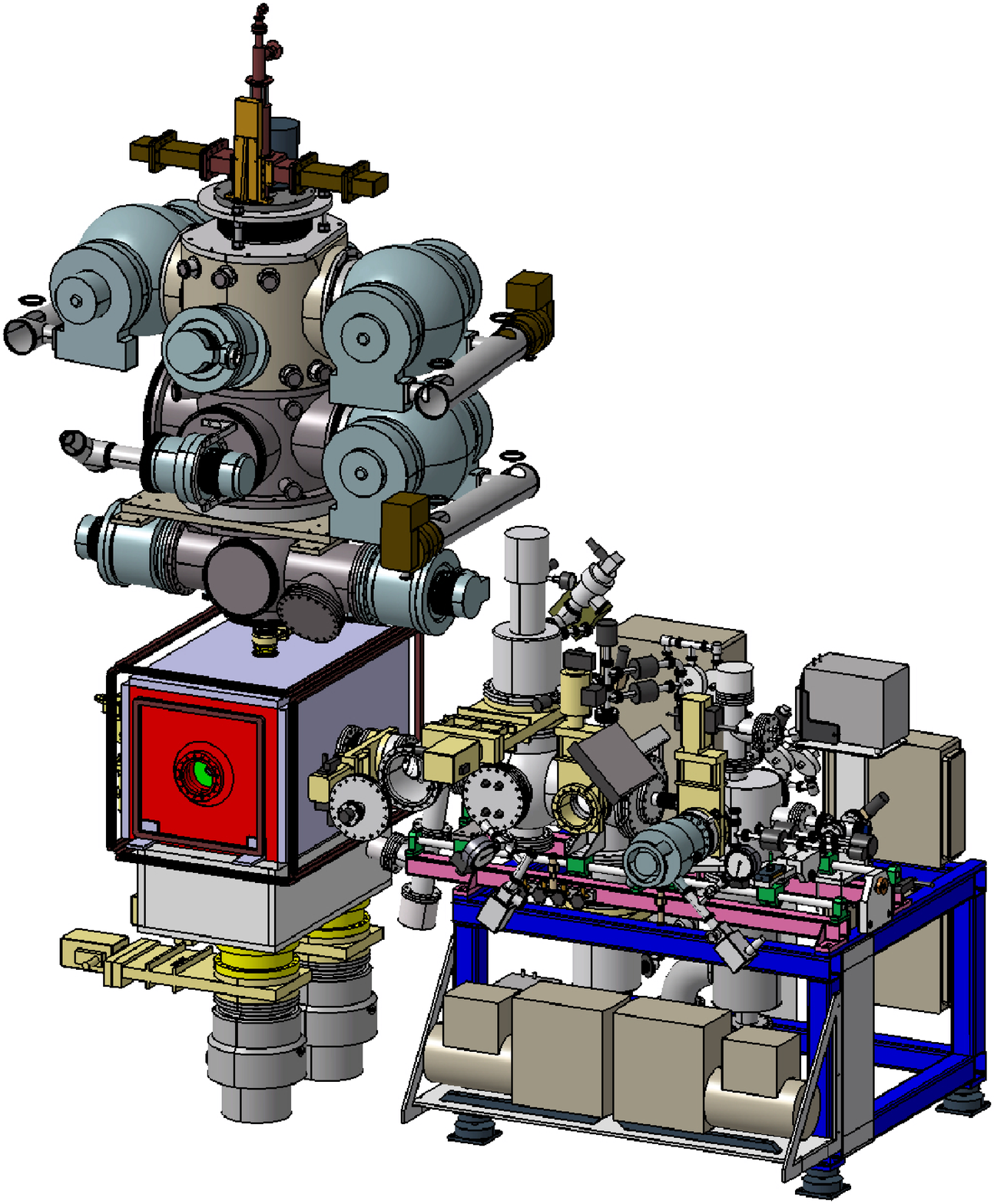}
\hspace{0.2cm}
\includegraphics[width=0.48\linewidth]{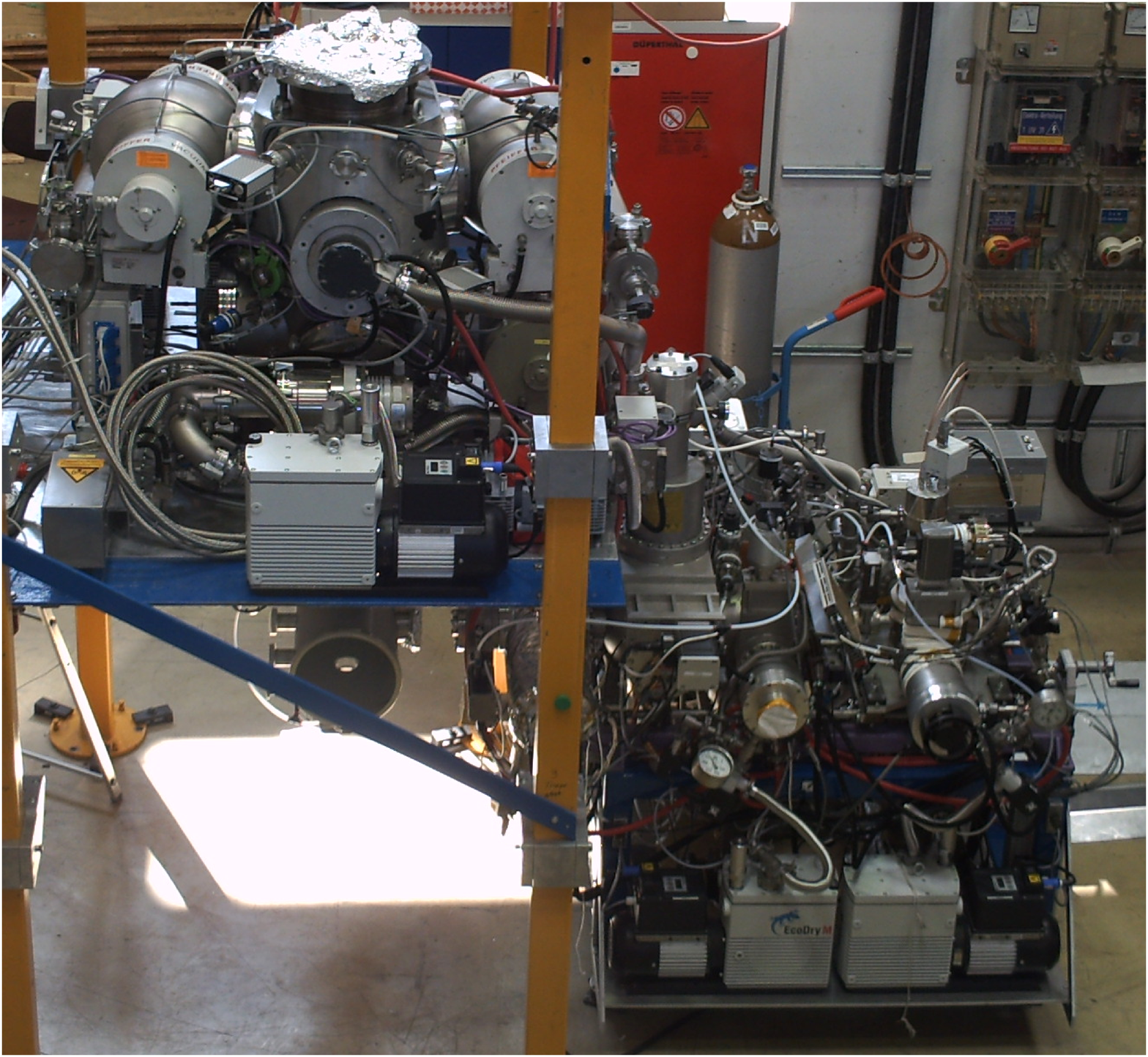}
  \parbox{14cm}{\caption{\label{fig:target}\small Left panel: The PAX target at the AD. The atomic beam source (ABS) is mounted on top of the target chamber which houses the storage cell and the detector system. The antiproton beam passes through the target from behind. The Breit--Rabi polarimeter (BRP) on the right is fed by a small sample beam extracted from the storage cell. Right panel: The ABS and the BRP with a small test chamber set up in J\"ulich.
}}
\end{center}
\end{figure}

\begin{figure}[hbt]
 \begin{center}
\includegraphics[width=0.75\linewidth]{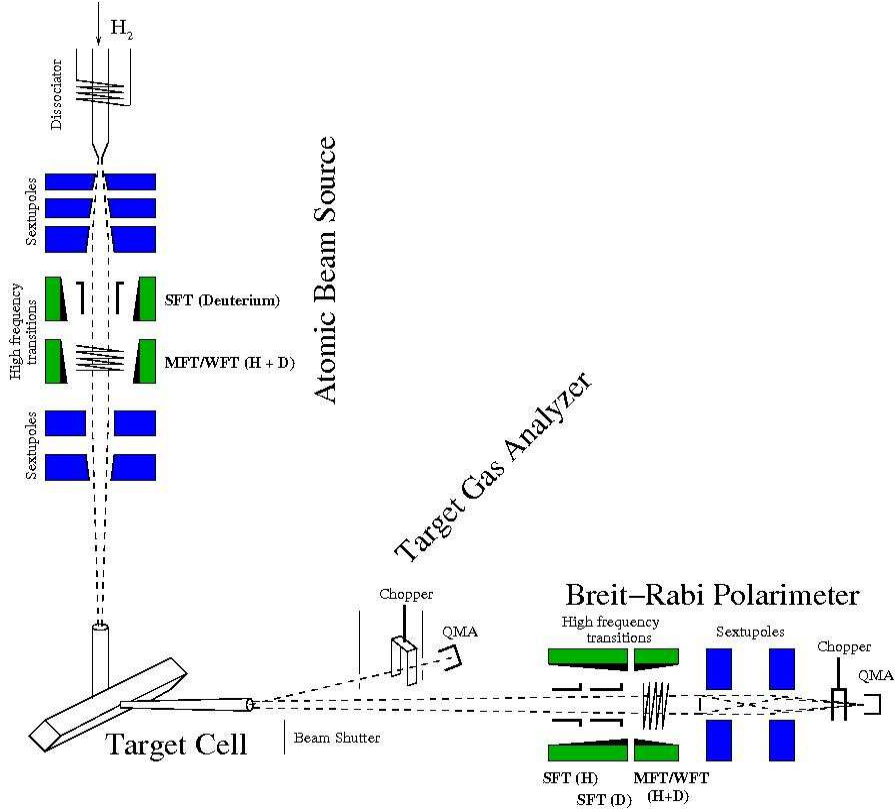}
  \parbox{14cm}{\caption{\label{fig:abs-brp-scheme}\small Schematic view of the target setup with the sextupole magnet systems and the radio-frequency transition units.
}}
\end{center}
\end{figure}

The PAX target arrangement is depicted in Fig.~\ref{fig:target}. It consists of an ABS, a storage cell, and a Breit--Rabi polarimeter (BRP). H or D atoms in a single hyperfine--state are prepared in the ABS and injected into the 
thin--walled storage cell. Inside the target chamber, the cell is surrounded by the detector system (see Sec.~\ref{sec:detector}). As shown in Fig.~\ref{fig:abs-brp-scheme}, a  small sample of the target gas propagates from the center of the cell into the BRP  where the atomic polarization is measured. Simultaneously the sampled gas enters the Target Gas Analyzer (TGA) where the ratio of atoms to molecules in the gas is determined. A weak magnetic holding field around the storage cell provides the quantization axis for the target atoms (coils indicated in brown on the edges of the target chamber in Fig.~\ref{fig:target}, left panel); it  can be oriented along the transverse ($x$), vertical ($y$), or longitudinal ($z$) direction.

With respect to the setup of the target used at HERMES, two major modifications have been implemented: 
\begin{enumerate}
 \item Fast switching of the target between operation with H and D  {\it without} changes to the hardware.
 \item Design and realization of an openable storage cell.
\end{enumerate}
Both aspects are addressed in detail in the following, including a short description of the main components of the target.

\subsubsection{Atomic beam source}
The atomic beam source consists of a dissociator, a powerful differential pumping system, a beam forming system, a sextupole magnet system to focus atoms with $m_S$ = +$\frac{1}{2}$ into the storage cell, and a set of adiabatic high--frequency transitions to manipulate the hyperfine population of the atomic beam. Injected fluxes of $\Phi^{ABS}$ $\approx$ $6.5 \times10^{16}$ atoms/s in case of hydrogen (2 hyperfine states injected) and $\Phi^{ABS}$ $\approx$ $5.8 \times10^{16}$ atoms/s in case of deuterium (3 states injected) have been observed. A schematic of the ABS is depicted in Fig.~\ref{fig:abs-brp-scheme}. A detailed description of the HERMES ABS and its performance can be found in ref.~\cite{hermes-abs}.

During the polarization buildup measurements for spin--filtering with antiprotons, the ABS will inject a vector polarized beam, both for H and D into the storage cell. For weak magnetic holding fields $B \ll B_c$ ($B_c^\mathrm{H}=507$~G for hydrogen, and  $B_c^\mathrm{D}=117$~G for deuterium) this can be accomplished  by the use of pure hyperfine states. The injected states together with the status of the HF--transition units are given in Table~\ref{tab:ABS_injection}.
\begin{table}[hbt]
\begin{center}
\begin{tabular}{|c|c|c|c|c|c|}
\hline
  & HFT               & inj. states & $P_e$ & $P_z$ & $P_{zz}$\\
\hline
H &  MFT 2--3          & $\ket{1}$   &  +1   & +1    &  --       \\
\hline
D &  SFT 2--6, MFT 3--4 & $\ket{1}$   &  +1   & +1    & +1      \\
\hline
\end{tabular}
\parbox{14cm}{\caption{\label{tab:ABS_injection}\small ABS injection modes during data taking. The table shows the  high frequency transition (HFT) units employed, the hyperfine states injected into the storage  cell and the resulting electron ($P_e$), nuclear ($P_z$) and tensor ($P_{zz}$) polarizations in the ideal case of 100\,\% efficiency of the sextupole system and transition units.
}}
\end{center}
\end{table}

\subsubsection{Storage cell}
\begin{figure}[hbt]
 \begin{center}
\includegraphics[width=0.49\linewidth]{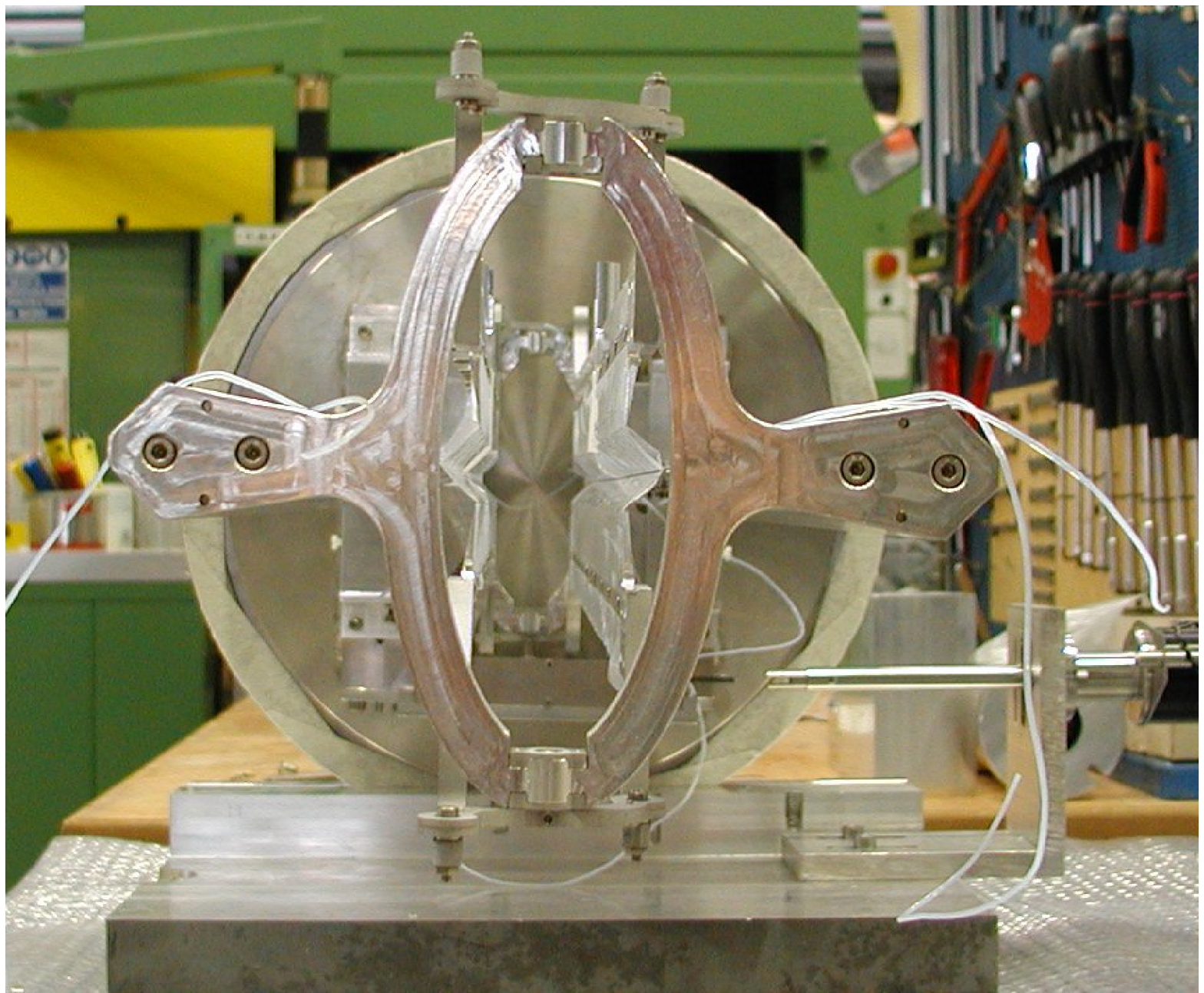}
\includegraphics[width=0.49\linewidth]{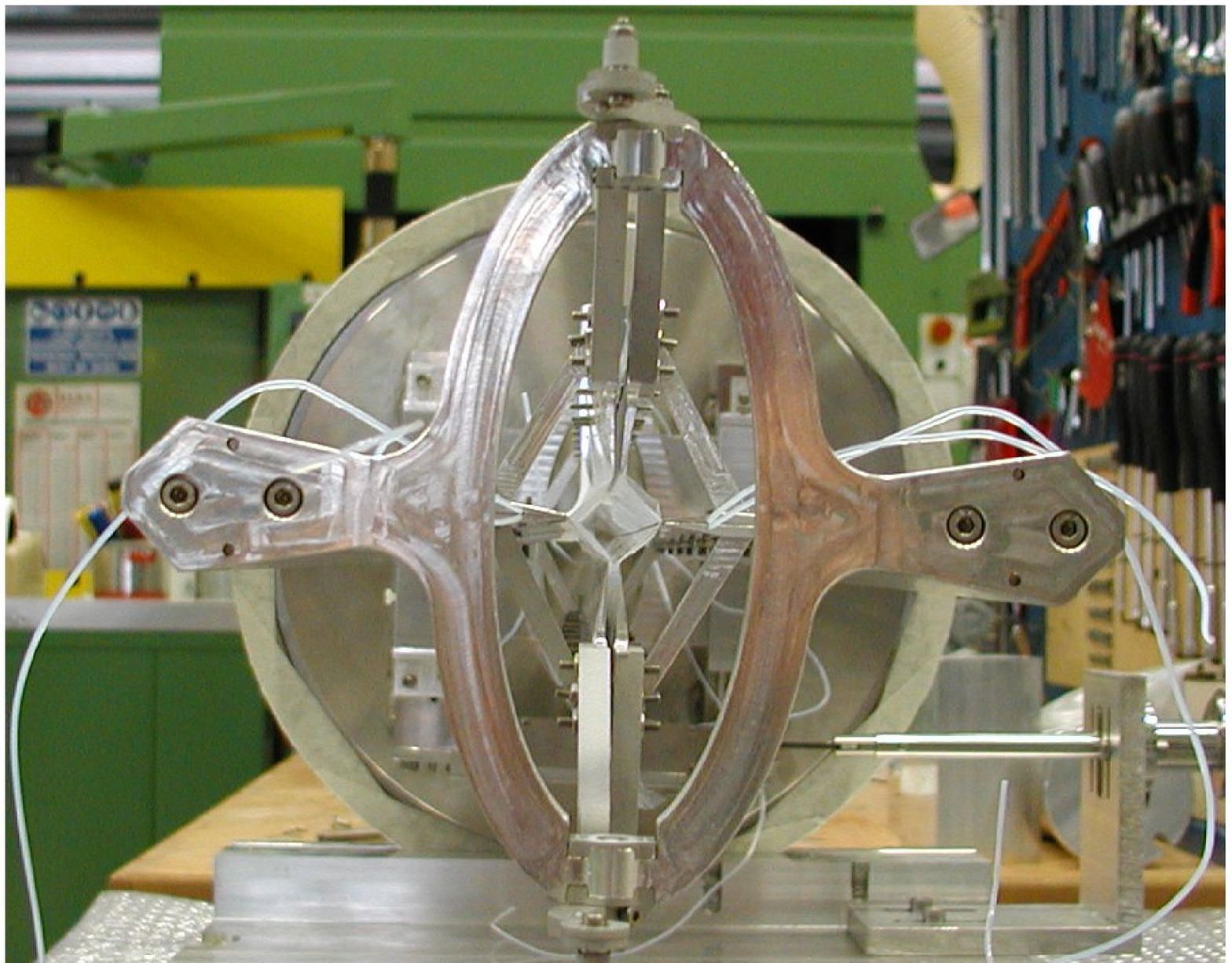}
  \parbox{14cm}{\caption{\label{fig:cell}\small View along the beam direction of the prototype of the openable storage cell with the thin Teflon walls when opened (left panel) and closed (right). A CAD view of the mechanics of the openable storage cell together with the detector system is shown in Figs.~\ref{fig:kammer-innen} and \ref{fig:DetectorLayout} (right panel).
}}
\end{center}
\end{figure}

\begin{figure}[hbt]
\begin{center}
\includegraphics[width=0.75\textwidth]{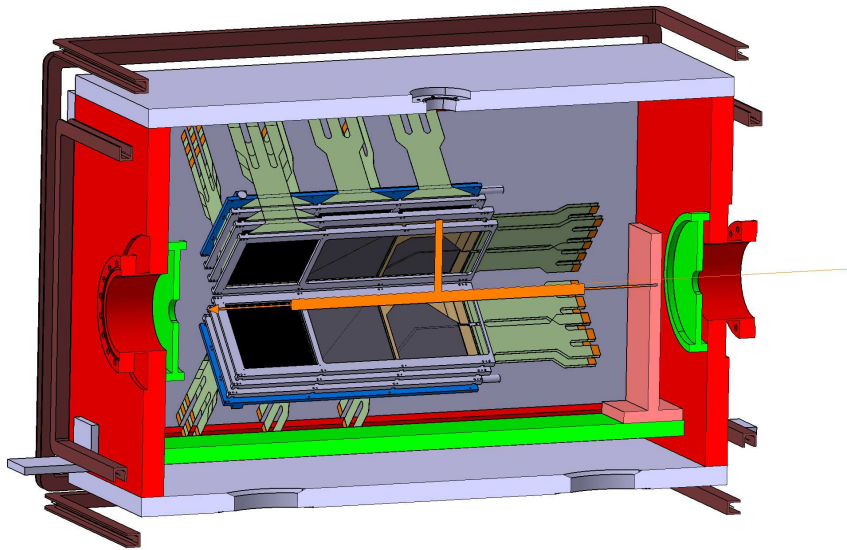}
 \parbox{14cm}{\caption{\label{fig:kammer-innen}\small CAD view into the target chamber, the beam passes through from right to left. The detector system is composed of up to in total 36 double--sided silicon strip detectors  around the 400~mm long storage cell target. (Only half of them are shown here. The third detector layer is an upgrade option.) Outside of the target chamber, the three pairs of Helmholtz coils are indicated (brown). The movable collimator system used to restrict the gas flow from the target chamber into the adjacent chambers up-- and downstream is shown in green.}} 
\end{center}
\end{figure}

As shown in Fig.~\ref{fig:envelope}, the beam sizes at injection before cooling are much larger than the anticipated 10~mm diameter of the storage cell. In order to be compatible with the antiproton beam operation at the AD, a dedicated storage cell has been developed that can be opened and closed at different times during the AD cycle. In particular at injection energy, a free space of 100~mm diameter is required at the cell position. As a compromise between density and acceptance angle, the closed cell has a square cross section of $10\times 10$~mm$^2$ and a length of 400~mm.  In order to allow for the detection of low--energy recoil particles, the cell walls are made from  a thin (5~$\mu$m) Teflon foil, a technique developed for the same reason for experiments at the Indiana Cooler~\cite{dezarn}. A prototype of the cell has been designed and produced in the mechanical workshop of Ferrara University, where also the target cell for the HERMES experiment~\cite{hermes-cell} was built. At this moment, a test of the cell is carried out  at J\"ulich. Photographs of closed and open cell are shown in Fig.~\ref{fig:cell}. A CAD view of the storage cell arrangement inside the target chamber is also shown in Figs.~\ref{fig:kammer-innen} and \ref{fig:DetectorLayout}.


\subsubsection{Target Gas Analyzer and Breit--Rabi Polarimeter\label{sec:tga-brp}}
\begin{figure}[t]
\begin{center}
\includegraphics[width=0.75\textwidth]{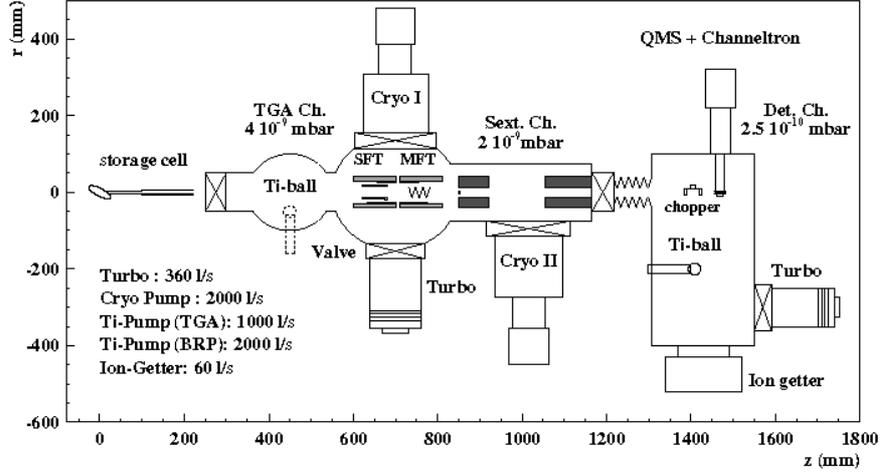} 
  \parbox{14cm}{\caption{\label{fig:BRP-vac}\small Schematic view of the BRP/TGA vacuum system. The sample beam exits the storage cell and travels from left to right. }}
\end{center}
\end{figure}

Although the recombination of target atoms into molecules on the cell surface is small, an accurate determination of the total target polarization based on the measured polarization of the atoms using the the BRP takes into account also the molecular fraction. The target gas analyzer (TGA) measures the atomic and molecular content of the gas extracted from the storage cell through the sample tube. The TGA arrangement consists of a chopper, a 90$^{\circ}$ off--axis quadrupole mass spectrometer (QMS) with a cross beam ionizer and a channel electron multiplier (CEM) for single ion detection. The TGA is integrated into the sextupole chamber of the BRP, which is pumped by two cryopumps and a titanium sublimation pump, with a total pumping speed of about 7000\,$\mathrm{\ell\,s^{-1}}$.  During operation, the pressure in the TGA detector is about 4$\cdot$10$^{-9}\,$mbar. Prior to normal operation, the TGA vacuum chamber is baked at temperatures up to 180$\,^{\circ}$C for 48 hours. The BRP/TGA vacuum scheme is shown in Fig.~\ref{fig:BRP-vac}. 
The TGA is mounted 7$^{\circ}$ off--axis with respect to the BRP, in order not to interfere with the beam entering the polarimeter. A chopper rotating  at a frequency of 5.5~Hz is periodically blocking the sample beam in order to allow subtraction of the residual gas signal. Particles entering the detector are ionized by electrons, mass filtered with the QMS, and finally detected by the CEM. A detailed description of the hardware and working principle of the HERMES Target Gas  Analyzer can be found in ref.~\cite{hermes-TGA}.

The Breit--Rabi polarimeter (BRP)  measures the relative populations $n_i$ of the hyperfine states of hydrogen (or deuterium) atoms contained  in the sample beam. From this measurement the absolute atomic polarizations can be calculated by applying the known field strength at the target. A schematic view of the BRP is shown in Fig.~\ref{fig:abs-brp-scheme}. From left to right, the sample beam leaves the sample tube of the target cell, encountering first two hyperfine transition units, then a sextupole magnet system and eventually the detector stage. During operation a differential pumping system keeps the pressure at  2$\times$10$^{-9}\,$mbar in the sextupole chamber and at  2.5$\times$10$^{-10}\,$mbar in the detector chamber.
Two transition units are used to exchange the populations between pairs of hyperfine states: a strong field transition unit (SFT) with tilted resonator that can be tuned for both $\pi$ and $\sigma$ transitions, and a medium field transition (MFT) unit that can induce various $\pi$  transitions according to the static field strength and gradient setting used. The sextupole system is composed of two magnets and spinfilters the sampled beam from the storage cell by focusing atoms with $m_s=+\frac{1}{2}$ towards the  geometrical axis of the BRP and defocusing atoms with $m_s=-\frac{1}{2}$. A beam blocker of 9~mm diameter  placed in front of the first sextupole magnet ensures that no atoms in $m_s = -\frac{1}{2}$ states can reach the detector. The detector stage is identical to the one employed for the target gas analyzer: a cross beam ionizer, a quadrupole mass spectrometer (QMS) and a channel electron multiplier (CEM). In contrast to the TGA, only hydrogen (or deuterium) {\it{atoms}} are detected by the BRP.  A detailed description of the hardware and working principle of the HERMES Breit--Rabi polarimeter is given in ref.~\cite{hermes-brp}.

We plan to use the Breit--Rabi polarimeter  both for polarized target operation with hydrogen and deuterium: 
\begin{itemize}
 \item{\bf Hydrogen operation:} 
      For the spin--filtering measurement with hydrogen the target polarization can be determined  by making use of the known $\bar{p}p$ analyzing power. At other energies where no analyzing powers are available, the BRP will provide the calibration standard by determining the target polarization.  In addition, the measurement of the target polarization with the BRP will serve as an important monitor of the ABS and the performance of the storage cell target. 
 \item{\bf Deuterium operation:} 
      Spin filtering an antiproton beam with a vector polarized deuterium target may well prove to be more efficient than using hydrogen. In view of the absence of measured  $\bar{p}d$ polarization observables, the BRP will provide during the AD studies the \textit{only} calibration standard to convert measured asymmetries into polarization observables. This entails also the determination of the target polarization during spin--filtering measurements with polarized deuterium. One option for the measurement of the resulting beam polarization is to operate the target during the spin--filtering measurement with deuterium and then inject unpolarized hydrogen into the storage  cell to make use of the known $\bar{p}p$ analyzing powers. The first spin--observables for $\bar{p}d$  can already be measured after a beam polarization has been produced through spin filtering, when under these conditions, the polarized target is operated with deuterium (see also Sec.~\ref{sec:polarimetry}).
\end{itemize}

\subsubsection{Dedicated modifications for the PAX experiment}
The BRP employs two sets of rf--resonators, one for hydrogen, and one for deuterium operation. The frequency for two--level transitions ($\Delta F = \pm 1$) is  of the order of the hyperfine splitting energy ($\Delta$W) at $B = 0$. For H it corresponds to the transition $F = 0 \rightarrow F = 1$ at $B = 0$, the famous 21~cm--line in the cosmic microwave background. The transition frequencies in hydrogen (deuterium)  $\Delta W/h$ are 1421.4~Mhz (327.4~MHz). 
 
The polarized target at the HERMES experiment has been operated for long periods with one target species only (H or D). In order to switch from H to D or vice versa, the resonating structures had to be exchanged, and a tedious tuning procedure had to be performed. The studies at the AD require to operate the BRP consecutively with H and D (or vice versa) with very short time intervals in between. Therefore, a system of resonators is required which can operate at frequencies for H and D without modifications in the setup. In addition, the magnetic component of the rf--field should be tilted by 45$^\circ$ with respect to the static field in order to allow both for $\sigma$ and $\pi$ transitions ($\Delta m_F$ = 0 and $\pm1$, respectively). If a single cavity can serve for both H and D operation, the existing hardware could be used, like static field magnets and the vacuum system. For the above mentioned reasons, a dual cavity for the PAX target polarimeter has been built which provides rf--field configurations for two--level transitions, both for hydrogen and deuterium in a tilted--field geometry. In particular, two independent pairs of resonator rods with separate coupling and pick--up loops are arranged parallel to the beam axis in two planes tilted by $\pm 45^\circ$ with respect to the median plane. It has been demonstrated that they can be tuned independently to their respective transition frequencies at about 1430~MHz (H) and 330~MHz (D) without interference. The details of the development can be found in ref.~\cite{pax-note-1-2009}.

\subsubsection{Target chamber and vacuum system}
\begin{figure}[hbt]
\begin{center}  
\includegraphics[width=0.75\linewidth]{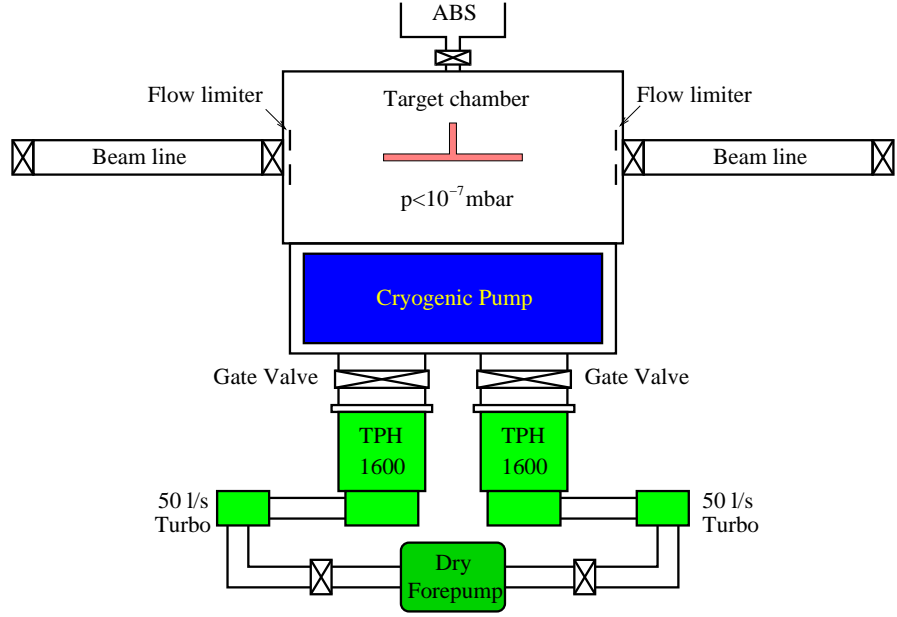}
 \parbox{14cm}{\caption{\label{fig:vac}\small Vacuum system at the target chamber}} 
\end{center}
\end{figure}

The target chamber hosts the storage cell and the detector system (Fig.~\ref{fig:kammer-innen}). The target  section has to be equipped with a powerful differential pumping system, that is capable to maintain good vacuum conditions in the adjacent  sections of the AD. The target chamber can be separated from the up-- and downstream sections of the beam pipe by two fast gate valves capable to close within $\approx$15~ms.  The vacuum system of the target chamber includes two turbo molecular pumps with 1600~$\ell$/s pumping speed, backed by smaller turbo molecular pumps and a dry forevacuum pump.  In addition, a large cryogenic pump with a nominal pumping speed of about 20000~$\ell$/s, mounted directly below the target chamber, will ensure that most of the target gas exiting the storage cell is pumped away in the target chamber (see Fig.~\ref{fig:vac}). 

Flow limiters installed between the target chamber and the  adjacent up-- and downstream beam pipes reduce the gas load into these sections. These beam pipes  will be coated with Non--Evaporable Getter (NEG) material to ensure a pumping speed per section of $\approx 3000$~$\ell$/s. It is necessary to activate and regenerate the NEG material at temperatures of 230--250$^\circ$C. Assuming an atomic beam intensity of $6 \times 10^{16}$~atoms/s (two states injected) and a pumping speed of 20000~$\ell$/s in the target chamber, the expected chamber pressure during target operation would  be about 1.7$\times 10^{-8}$~mbar.  With flow limiters of 30~mm diameter at the exit and entrance of the target chamber (see Fig.~\ref{fig:kammer-innen}), the total flux from the storage cell and the target chamber into the adjacent beam pipes will correspond to  about $2 \times 10^{14}$~atoms/s, resulting in a pressure in the $10^{-9}$~mbar range in the up-- and downstream sections.

\subsubsection{Unpolarized gas feed system\label{sec:ugfs}}
One option to measure the beam polarization exploits the analyzing power of elastic $\bar{p}\vec{p}$ scattering (see also Sec.~\ref{sec:polarimetry}). After spin--filtering with polarized H or D targets, it is possible to interrupt the atomic beam from the source and to inject  unpolarized H$_2$ (or any other gas species) into the storage cell. To this end, an Unpolarized Gas Feed System (UGFS) is used to inject  calibrated gas fluxes into the storage cell through a separate feeding tube. 
The UGFS 
is capable to produce gas flows in the range of $10^{14}$ to $10^{18}$~H$_2$/s. 

\subsubsection{Target holding field and compensation coils}
The target holding fields are generated by  racetrack coils mounted on the outside of the target chamber (see Fig.~\ref{fig:target} (left panel), and Fig.~\ref{fig:kammer-innen}). Three pairs of coils provide a holding field in $x$, $y$, and $z$ direction. The coils generating the $x$ and $y$ (transversal) fields carry about 750 Ampere$\times$turns (cross section $13\times13$~mm$^2$), while the longitudinal coils carry about 1125 Ampere$\times$turns. The average field generated by the transverse coils at the cell is 1.18~mT and almost constant, while the longitudinal field varies from 1.0 to 1.5~mT. The integrated field in beam direction is $0.77\times$10$^{-3}$~Tm and $2.81 \times$10$^{-3}$~Tm, respectively for the transversal and the longitudinal case. The transverse field causes a transverse displacement of the stored beam which will be compensated by short magnets placed on the beam pipe in front and behind the target chamber. For a  beam momentum of 300~MeV/c, the angle kick at the target due to the transverse holding field (without compensation)   corresponds to  $<1$~mrad.

\subsubsection{Vacuum interlock and safety}
One gate valve per large turbomolecular pump will isolate the pumps from the target chamber in case of necessity.  If one of the two turbomolecular pumps fail, the corresponding gate valve will close immediately. The forevacuum pump keeps the vacuum below the gate valve to reduce leakage through the gate valve. The complete low--$\beta$ insertion can be sealed off from the rest of the AD by two gate valves (shown in Fig.~\ref{fig:setup-phase1}).

All main ports at the target chamber will be equipped with fast  shutters capable to close within 15~ms, {\it i.e.} the up-- and downstream ports at the target chamber towards the AD, the valve connecting the ABS to the target chamber, and the valve towards the BRP.

\subsection{Detector system \label{sec:detector}}
The PAX collaboration is developing a detector system to efficiently determine the polarizations of beam (or target) by measuring the polarization  observables in $\bar{p}p$ elastic scattering. 

The detector has been designed to meet the following requirements:
\begin{itemize}
 \item provide a measurement of polarization observables in both $pp$ (at COSY~\cite{COSYLOI}) and            $\bar{p}p$ elastic scattering at AD~\cite{ADLOI};
 \item operate in vacuum to track low--momentum particles in the kinetic energy range from a few to a few tens of MeV;
 \item allow for the antiproton beam envelope at injection and allow for the storage cell to be opened         during injection at the AD; 
 \item provide large coverage of solid angle and luminous volume.
\end{itemize}

\begin{figure}[hbt]
\begin{center}
\includegraphics*[width=0.45\textwidth]{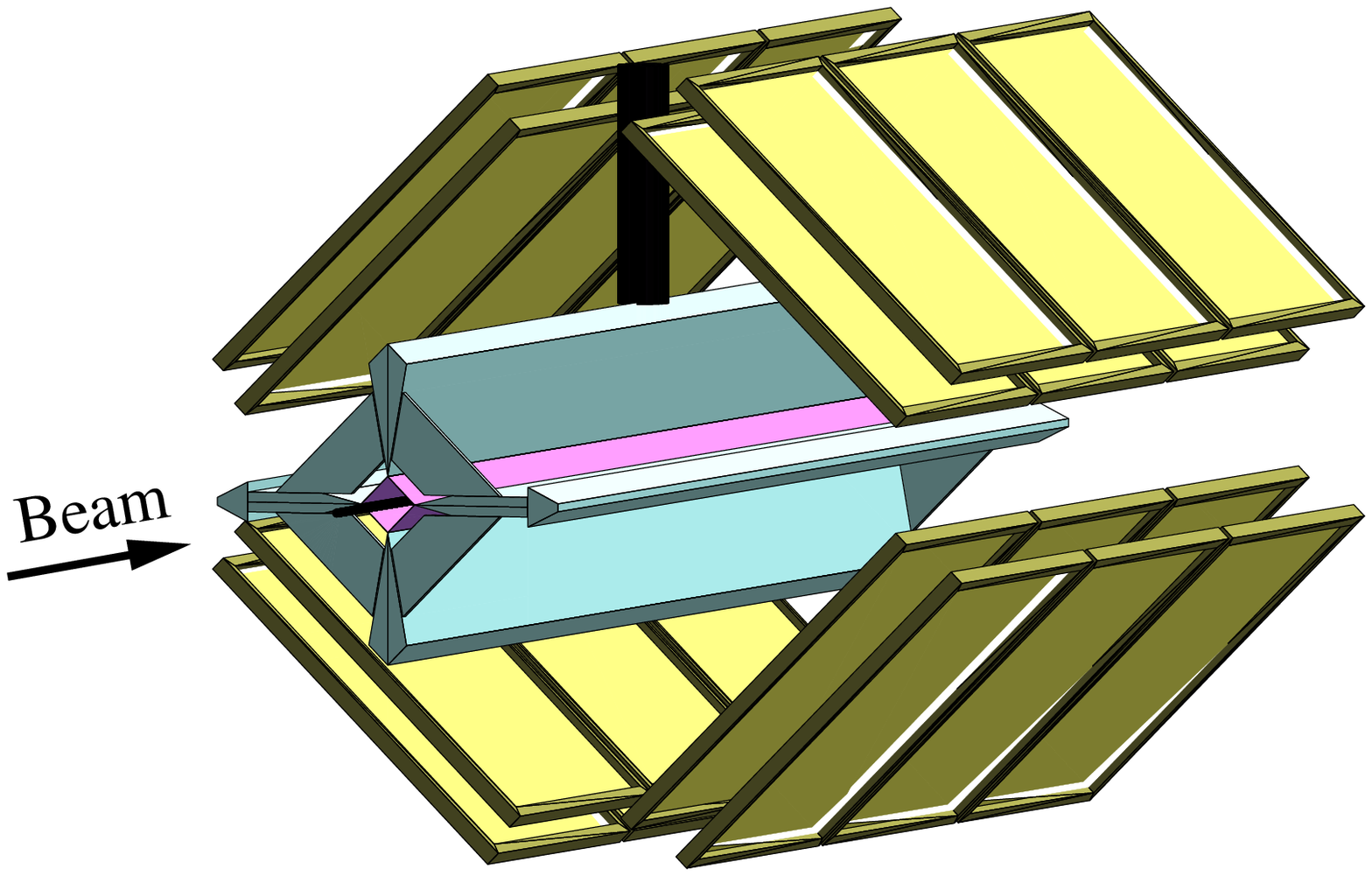}
\includegraphics*[width=0.53\textwidth]{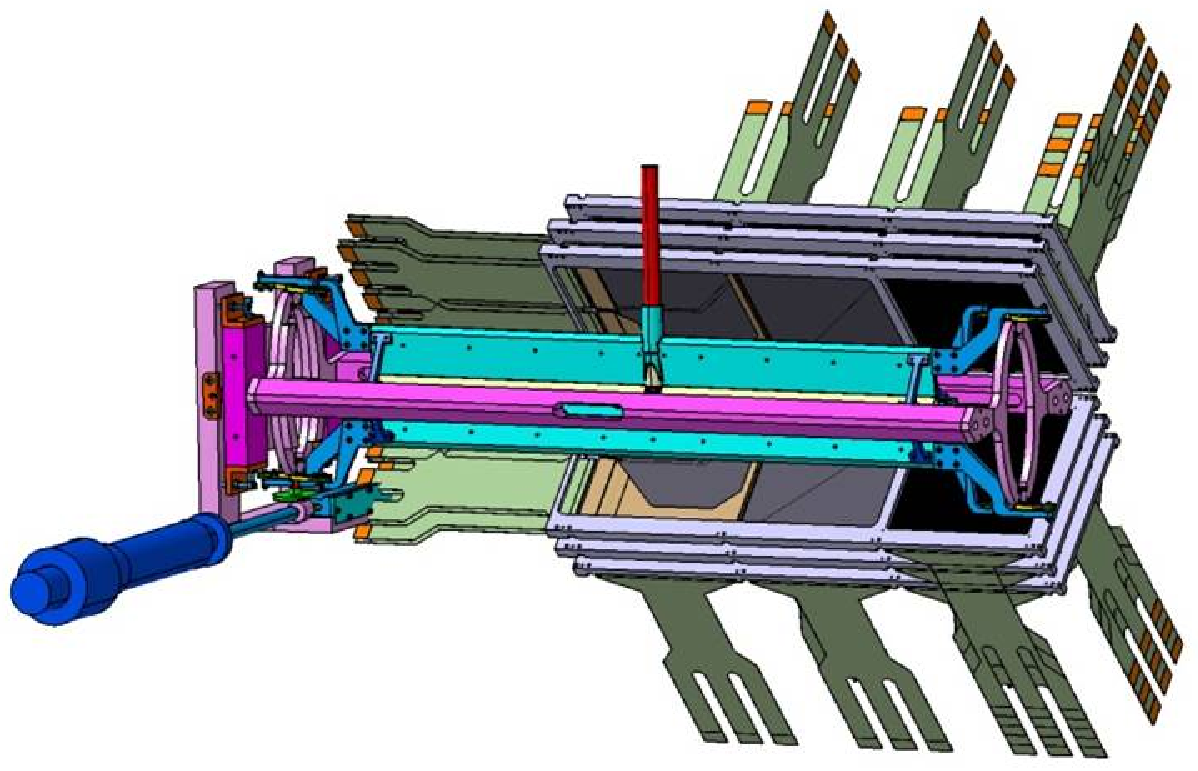}
 \parbox{14cm}{\caption{\label{fig:DetectorLayout}\small Left panel: Sketch of the silicon detector, drawn by the simulation package, surrounding the target cell. (The third detector layer is an upgrade option.) The Teflon cell walls are shown in violet, the aluminum target frame in sky blue, and the sensitive modules in yellow. Right panel: The complete system with openable storage cell support and gear, silicon detector frames, and kapton connection to the front--end electronics. (Only half of the detectors are shown.)}} 
\end{center}
\end{figure}

\subsubsection{Detector configuration}
The detector system is based on silicon microstrip detectors and the design follows closely the one recently developed at IKP~\cite{SPIN} for the ANKE experiment at J\"ulich. A detailed description of the ANKE detector system can be found in ref.~\cite{Mussgiller}.

The PAX detector is based on two layers of double--sided silicon--strip sensors  of large area ($97 \times 97$~mm$^2$) and standard thickness of 300 $\mu m$. Eventually, the number of silicon layers can be increased to three to provide redundancy of the obtained track information. A pitch of 0.76~mm provides the required vertex resolution of about 1~mm, while minimizing the number of channels to be read out. 

The detector setup, shown in Fig.~\ref{fig:DetectorLayout}, is placed around the 400~mm long and $10\times10$~mm$^2$ cross section of the storage  cell. Three adjacent detector layers along the beam direction  cover the central and forward sections of the storage cell in order to maximize the acceptance for elastic scattering. Events occur mainly in the central region of the cell where the target density has its maximum. The minimum radial distance of the inner silicon layer is defined by the space required for the injection tube and the movement of the cell at injection. The 10~mm radial distance between two silicon layers is chosen to maximize the azimuthal acceptance while preserving the resolution on the vertex.

\begin{figure}[hbt]
\begin{center}
\includegraphics[width=0.30\textwidth]{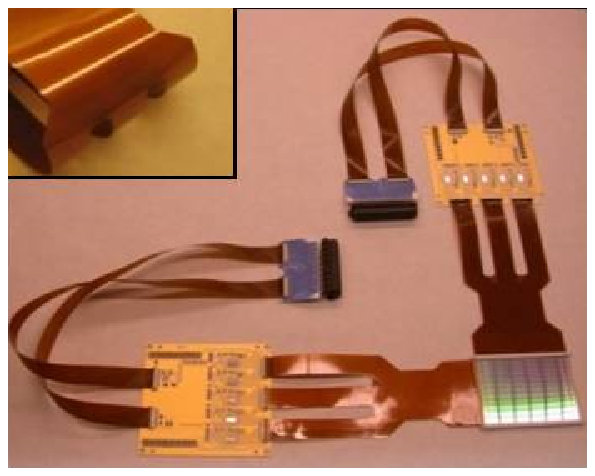}
\includegraphics[width=0.475\textwidth]{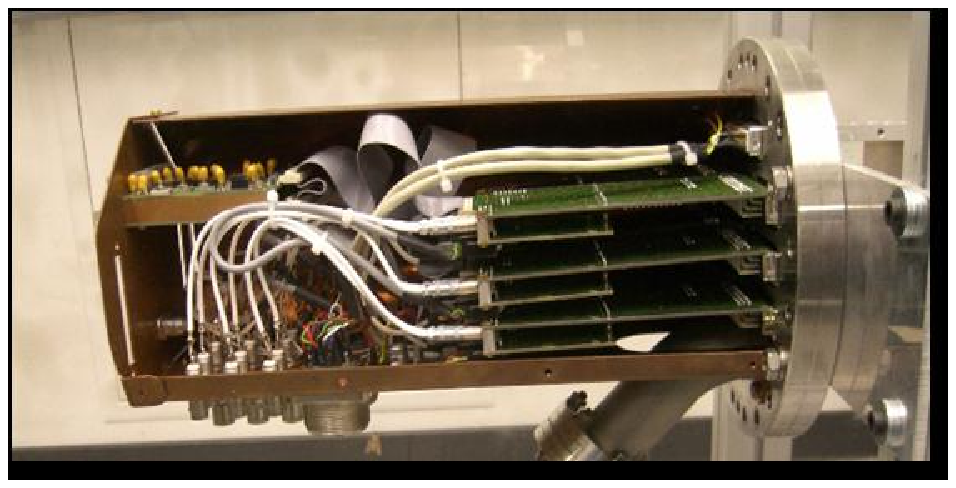}
\includegraphics[width=0.18\textwidth]{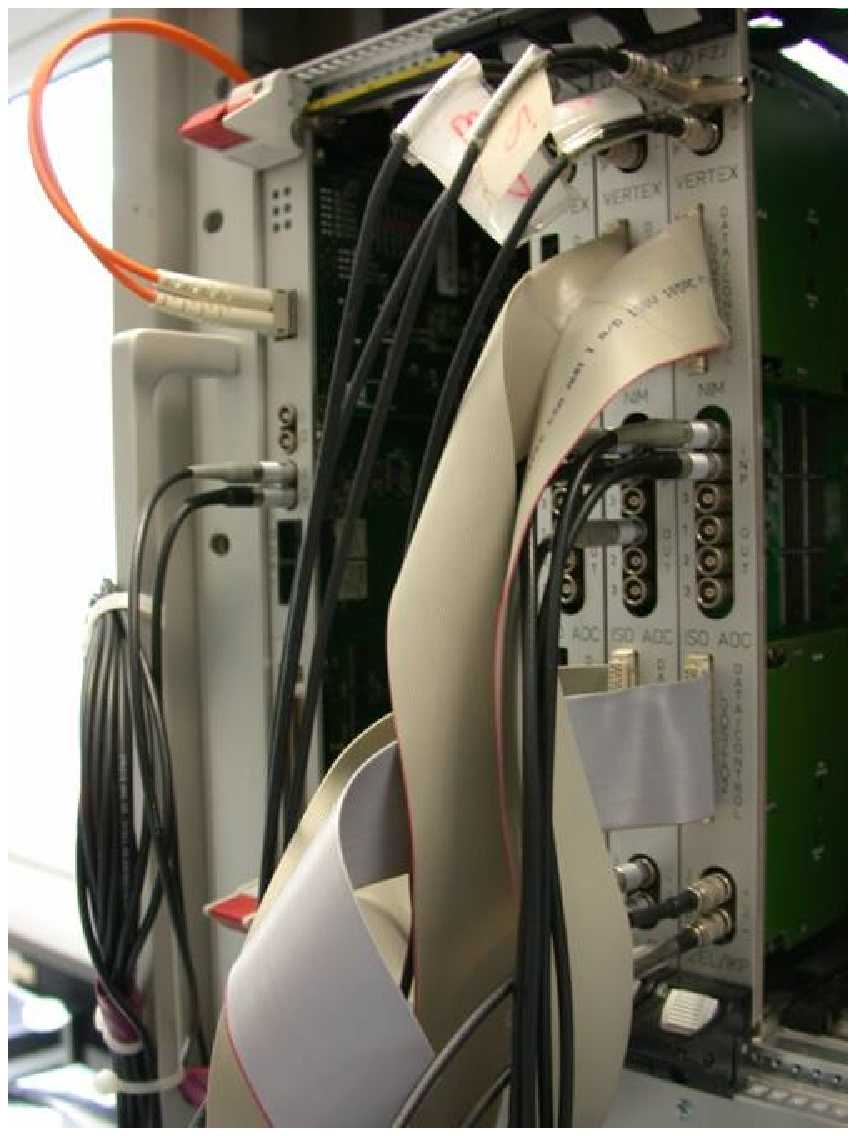}
 \parbox{14cm}{\caption{\label{fig:Electronic}\small The developed electronic chain: In--vacuum boards with 
self--triggering chips (left panel), front--end interface boards (center), and complete vertex board with 12 bit ADC and sequencer (right panel).}} 
\end{center}
\end{figure}

The read--out electronics (see Fig.~\ref{fig:Electronic}) are based on a scheme that has been developed for the ANKE detector at COSY. The in--vacuum board carries the read--out chips with 11 MeV linear range, and a time resolution of better than 1~ns, sufficient to  provide a fast signal for triggering. The interface card outside the vacuum provides power supplies, control signals, trigger pattern  threshold and calibration pulse amplitudes to the front--end chips. The vertex board developed at J\"ulich comprises a sequencer together with a 12 bit ADC with 10~MHz sampling; it allows common--mode correction for hardware zero--suppression to reduce the output flow to 0.1~MByte/s with less than 50 $\mu$s dead--time. A programmable trigger and a prescaler board have been developed to provide a flexible trigger logic.

\subsubsection{Detector simulations}
For the detector optimization and performance estimate, {\it i.e.} regarding the uncertainty in the measured kinematical parameters and the background rejection capability, the original software based on the GEANT4~\cite{Geant4} package was used. 

\begin{figure}[t]
\begin{center}
\includegraphics[width=0.29\textwidth]{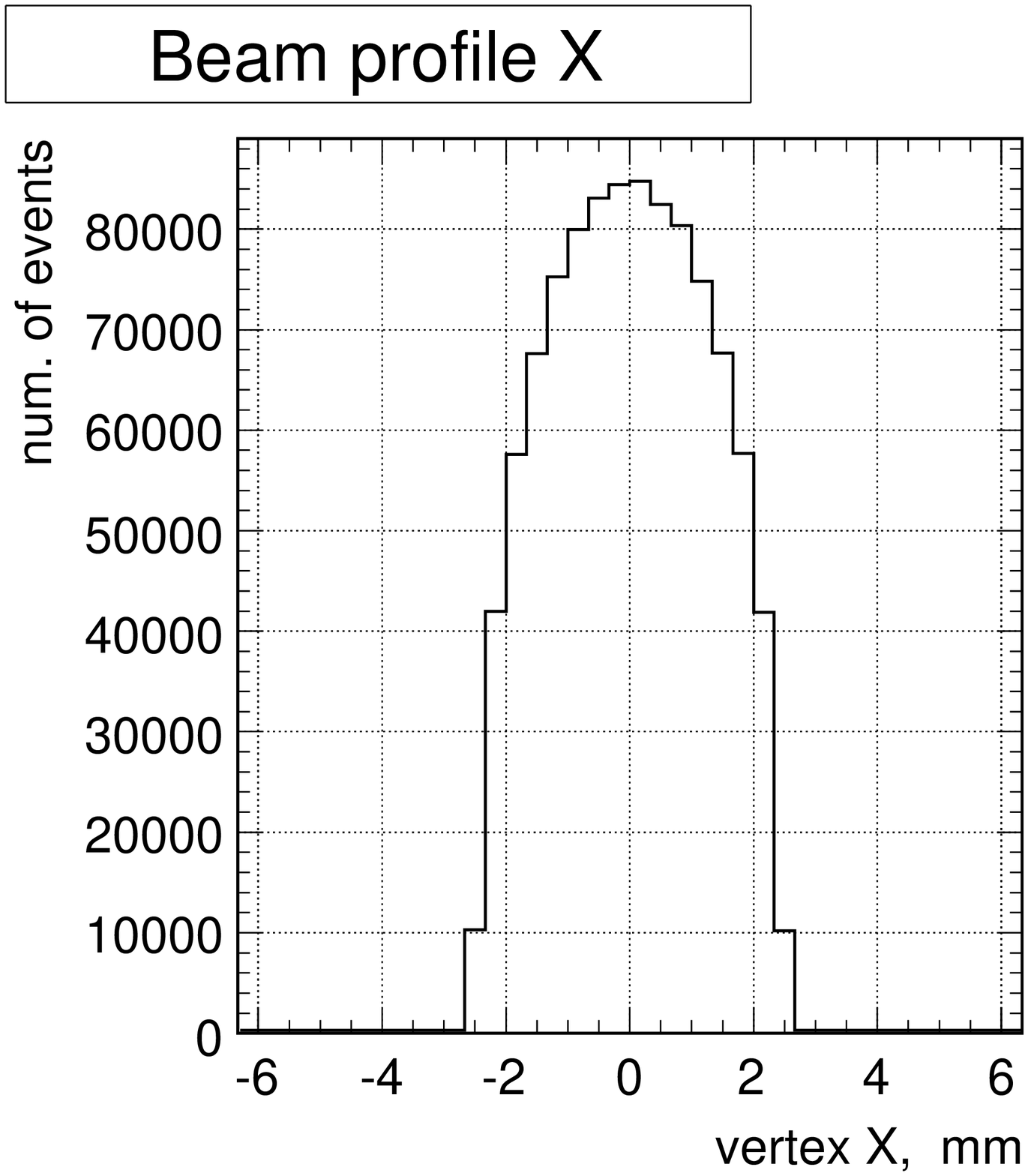}
\includegraphics[width=0.29\textwidth]{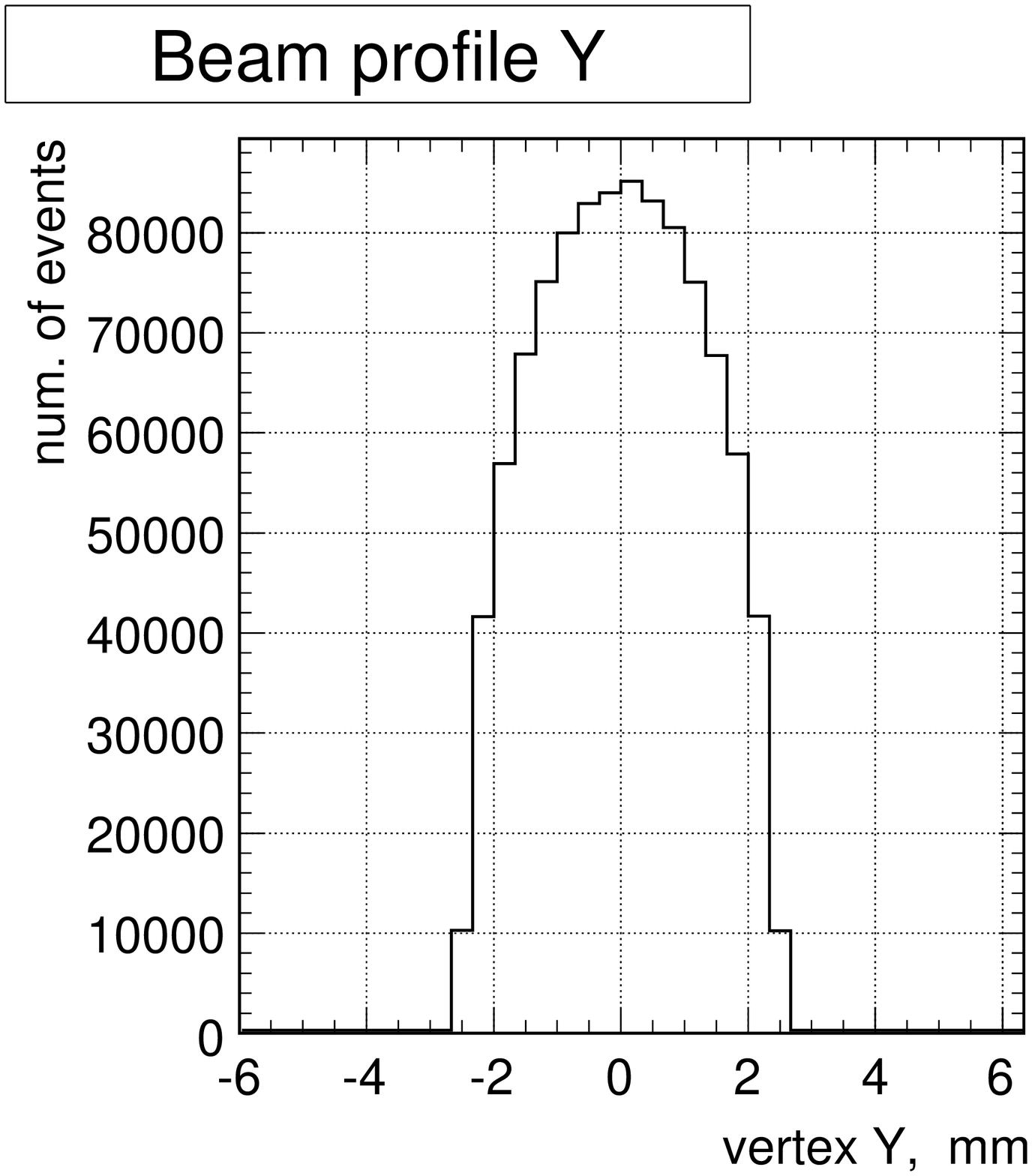}
\includegraphics[width=0.39\textwidth]{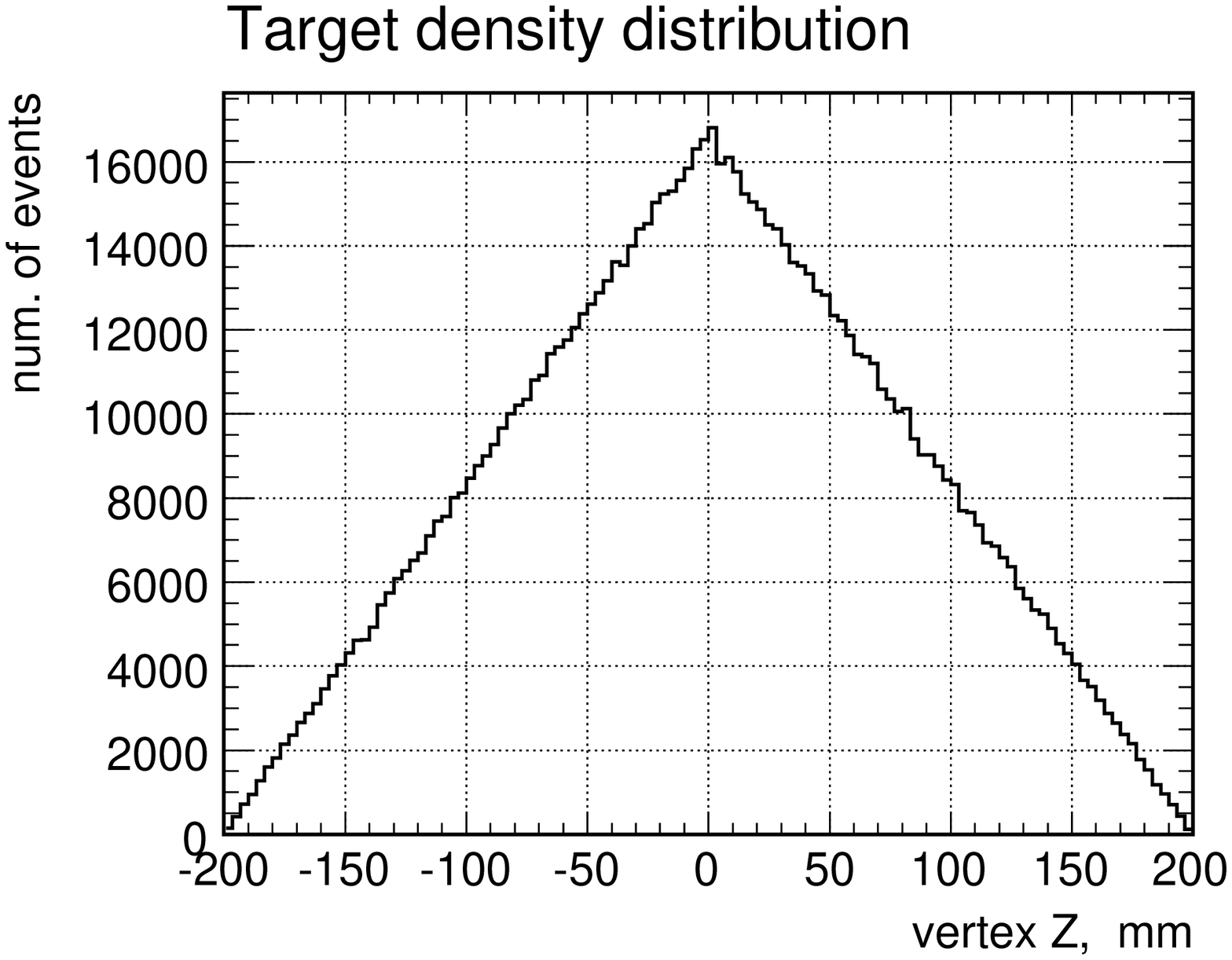} 
 \parbox{14cm}{\caption{\label{fig:Vertex}\small The spatial distribution of the generated $\bar{p}p$ interaction vertices in the target.}} 
\end{center}
\end{figure}

A 400~mm long target cell of $10\times10$~mm$^2$ cross-section and 10~$\mu$m thick Teflon walls is assumed. The simulation accounts for the cell frames and the injection tube for polarized gas from the atomic beam source (see Fig.~\ref{fig:DetectorLayout}). The description of the mechanical parts of the target and the corresponding drawings can be found in ref.~\cite{PaxInternal}. The vertex (interaction point) is randomly generated, depending on the target gas density distribution and the transverse beam size (see Fig.~\ref{fig:Vertex}).
Three beam energies (43, 120 and 220 MeV) have been chosen as representative for the anticipated measurements at the AD, as listed in Table~\ref{tab:BeamEnergies}. 
\begin{table}[hbt]
\begin{center}
\begin{tabular}{|l|r|r|r|r|}
\hline
Beam energy & MeV &  $43$ & $120$ & $220$ \\
Beam momentum  & MeV/c  & $286$ & $491$ & $679$ \\
$\sqrt{s}$ & MeV & $1897.9$ & $1935.6$ & $1983.5$ \\
$\sigma_{tot}$ & mb &  $250$ & $175$ & $145$ \\
$\bar{p}$ annihilation in material & \% of events &  $16.4$ & $6.7$ & $0.7$ \\
\hline
\end{tabular}
 \parbox{14cm}{\caption{\label{tab:BeamEnergies}\small Simulated beam energies and relevant parameters.}} 
\end{center}
\end{table}

In order to estimate the accepted event rate, the total cross section $\sigma_{tot}^{\bar{p}p}$ has been taken from ref.~\cite{klempt}: it ranges from 250~mb at 43 MeV to 145~mb at 220 MeV beam energy. The primary interaction of the $\bar{p}$ beam with the proton target is divided into three sub--processes: elastic, inelastic (annihilation), and charge-exchange. Each of these are separately simulated with relative intensities taken from ref.~\cite{klempt}.

\begin{enumerate}
     \item $\bar{p}p\to\bar{p}p$ elastic scattering (33\% of $\sigma_{tot}$) is based on  model predictions of the differential cross section $d\sigma/d\Omega$ and the analyzing power $A_{y}$~\cite{Mull,Hippchen,Haidenbauer};
     \item $\bar{p}p\to X$ inelastic interactions (60\%  of $\sigma_{tot}$) are simulated by the chirally invariant phase--space decay model (CHIPS)~\cite{Wellish1}, included in GEANT4; the laboratory momentum spectra of all secondaries are almost independent of the antiproton beam energy, since they are defined by the center--of--mass energy ($\surd s$) which is dominated by $\bar{p}$ and $p$ masses, see Table~\ref{tab:BeamEnergies}, thus the conditions for detector are the same for all energies. The inelastic interaction produces the primary background.
     \item $\bar{p}p\to \bar{n}n$ charge exchange scattering (7\%  of $\sigma_{tot}$) is based on the experimental data taken from~\cite{klempt} because the  CHIPS model does not agree with them.      Charge--exchange scattering is not expected to produce background, since most likely it does not   generate any signal in the silicon detector. Nevertheless, the events are accounted for in the simulation.
\end{enumerate}

Electromagnetic interactions of antiprotons with material (ionization losses and multiple scattering) do not differ from those for protons. Elastic and inelastic hadronic interactions of antiprotons with matter are generated using the parametrized interaction models of GEANT4. The CHIPS model~\cite{Wellish1} makes possible to simulate $\bar{p}$ annihilations in different materials. The model is well tested for $\bar{p}p$ annihilation interactions using two--particle final state branching ratios. The model is also used for annihilation at rest of some antiparticles $\pi^{-}$, $K^{-}$, $\Sigma^{-}$, $\overline{\Sigma^{-}}$, $\tau^{-}$, etc. 

\begin{figure}[hbt]
\begin{center}
 \includegraphics*[width=0.49\textwidth]{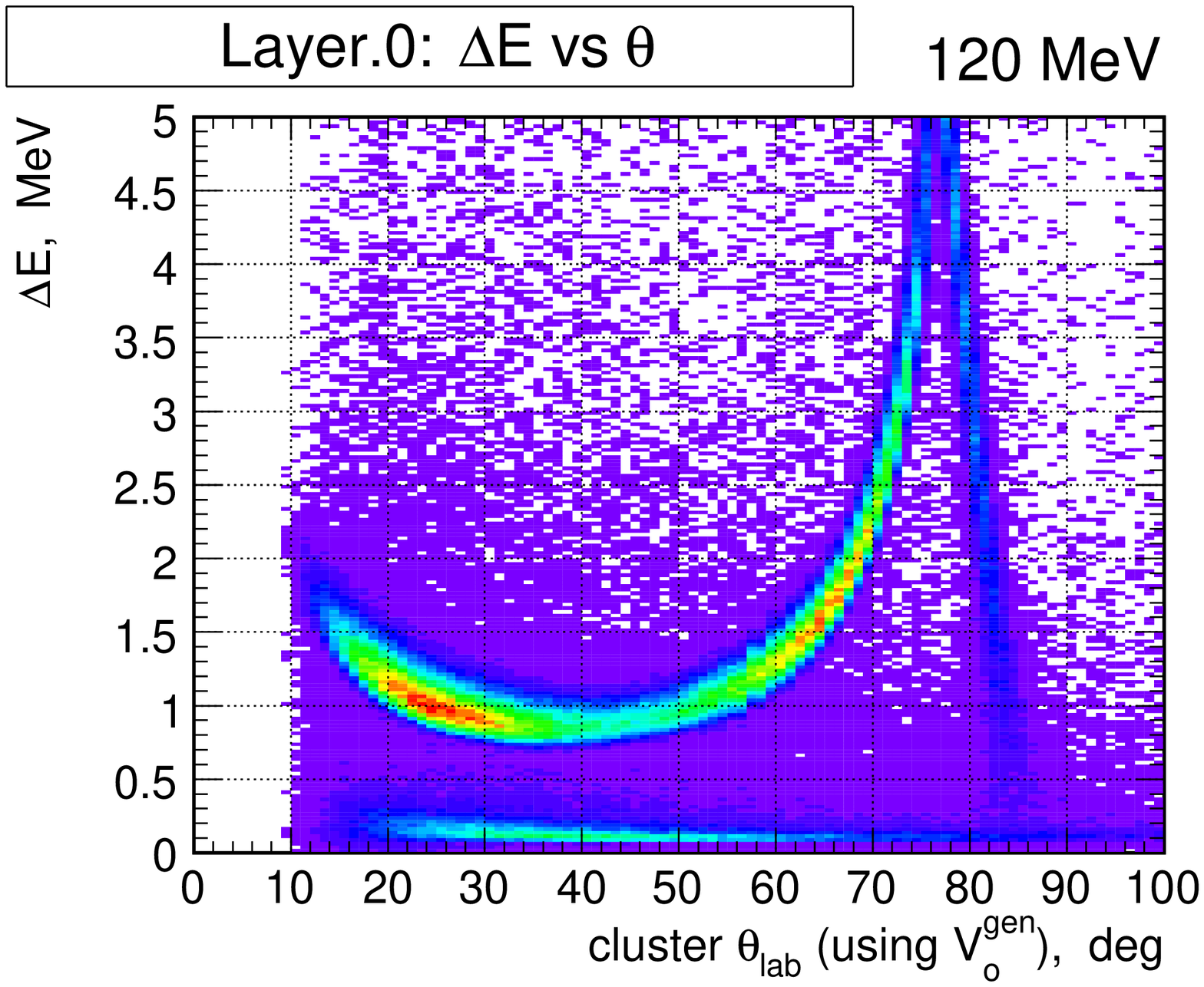}
\includegraphics*[width=0.49\textwidth]{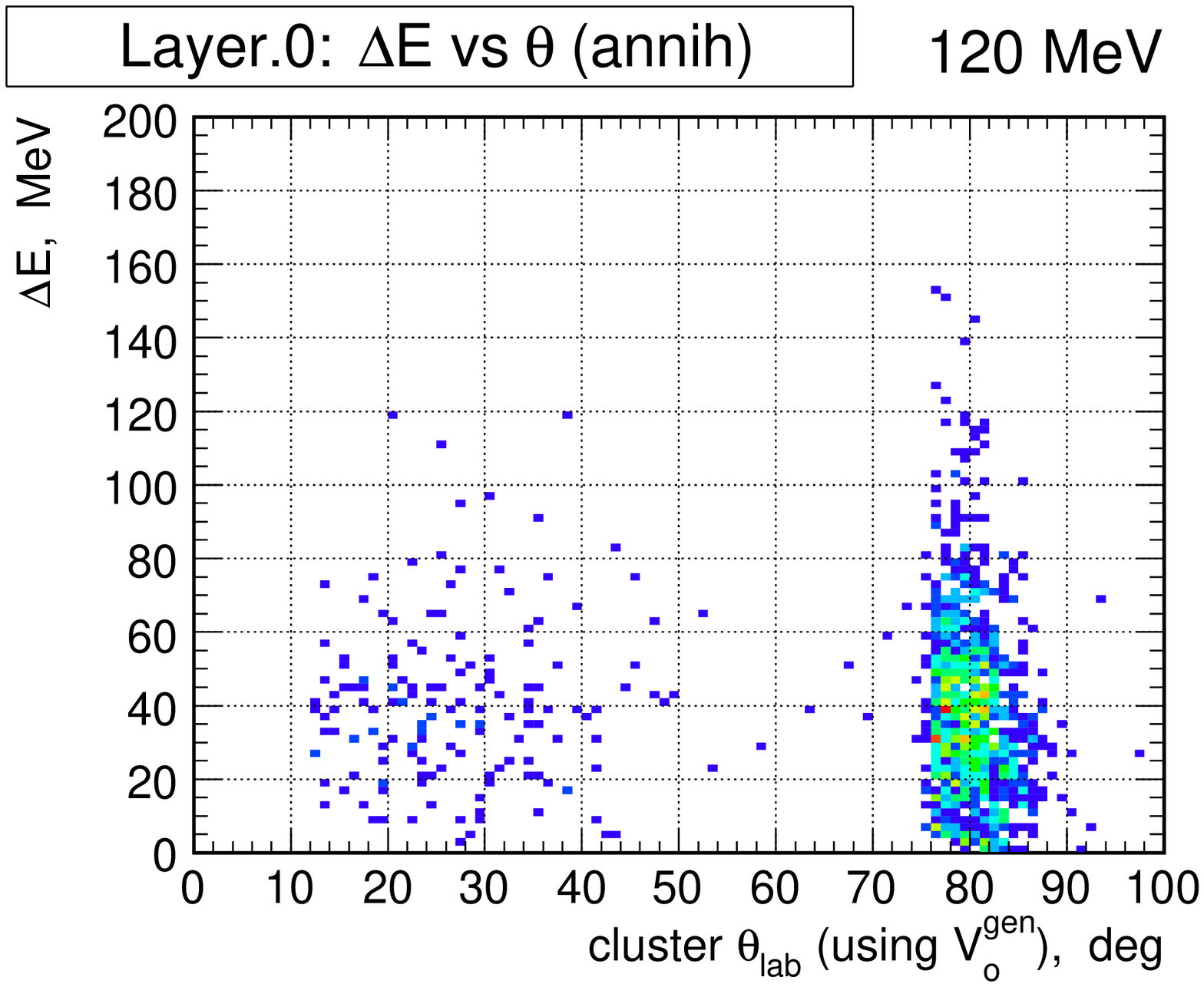}
 \parbox{14cm}{\caption{\label{fig:ClusterDe2Theta}\small Deposited energy as a function of the scattering angle $\vartheta_{lab}$ in the first silicon layer at 120 MeV beam energy, for non--annihilating particles (left) and for annihilating antiprotons (right). Signals below 0.5~MeV correspond to pion energy deposits, where pions stem from the annihilation of antiprotons in the material. Only primary elastic events are included.}} 
\end{center}
\end{figure}

In Fig.~\ref{fig:ClusterDe2Theta}, the energy depositions in the first silicon layer  is shown for 120 MeV beam energy as a function of the scattering angle in the laboratory frame $\vartheta_{lab}$. The response of the second layer is almost the same. In all calculations, the  intrinsic detector resolution on $\Delta E$ was considered to be much better than fluctuations. Signals below 0.5 MeV correspond to background energy deposits (mainly from  pions) and were removed. The angular range of stopped protons, between 80$^\circ$ and 90$^\circ$ (left panel), correspond to the angular range of antiproton annihilations (right panels). The number of stopped protons is two orders of magnitude larger than that of antiprotons and reflects the fact that antiprotons from the beam are mainly scattered in the forward direction. 

The over--constrained elastic kinematics allows event reconstruction based on the two scattering angles only. Energy information (precisely known only for stopped particles) is not required.  The deposited energies in the both layers at $\vartheta$ and $\phi$ make possible to reconstruct the particle kinetic energy. The most important parameters (kinematically independent) to describe the elastic scattering are $\phi_{n}$, the angle between the normal to the reaction plane and the $y$-axis, and $\vartheta_{lab}$, the scattering angle. Reliable and precise reconstruction of these parameters for each event makes it possible to measure any spin observable.

The reaction plane is reconstructed using 3D--coordinates of four silicon hits (two for each scattered track) with deposited energies above threshold. The reconstruction algorithm is based on the {\it orthogonal regression method} providing a parameter characterizing 'goodness of fit' ($\chi^2$) and has been developed in order to define the expected uncertainties caused by the multiple scattering and the energy losses.

From the measured count rates (or yields) at different beam and/or target polarizations, the beam or target polarization can be extracted using the available analyzing powers (or spin correlations, once measured) or vice versa. A powerful formalism to reconstruct  polarization observables from the measured yield in different spin--states of beam and target is the so--called {\it diagonal scaling} method~\cite{meyer2}.

\subsubsection{Detector performance: Acceptance and event rate estimate}
\begin{figure}[h]
\begin{center}
\includegraphics*[width=0.45\textwidth]{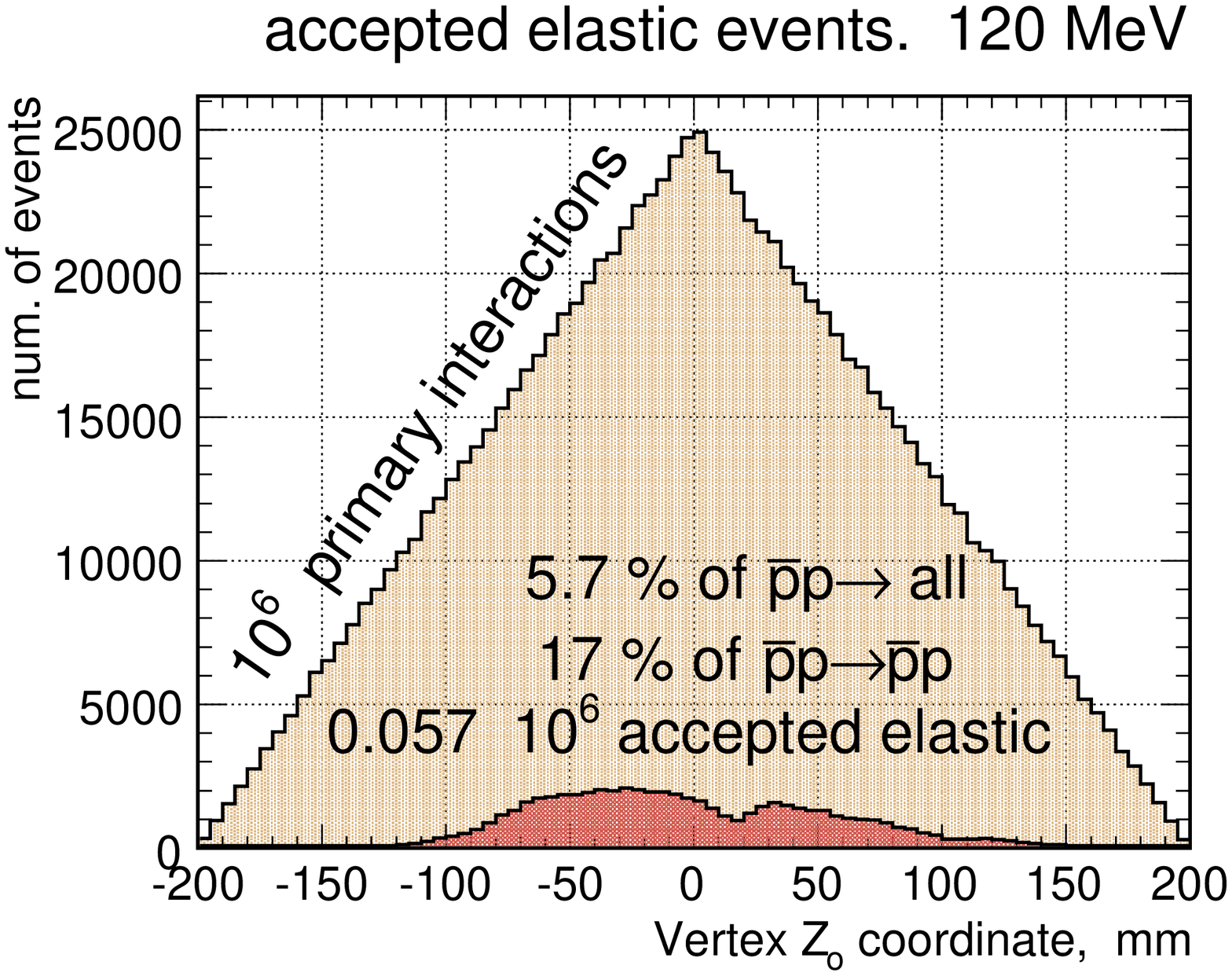}
\includegraphics*[width=0.45\textwidth]{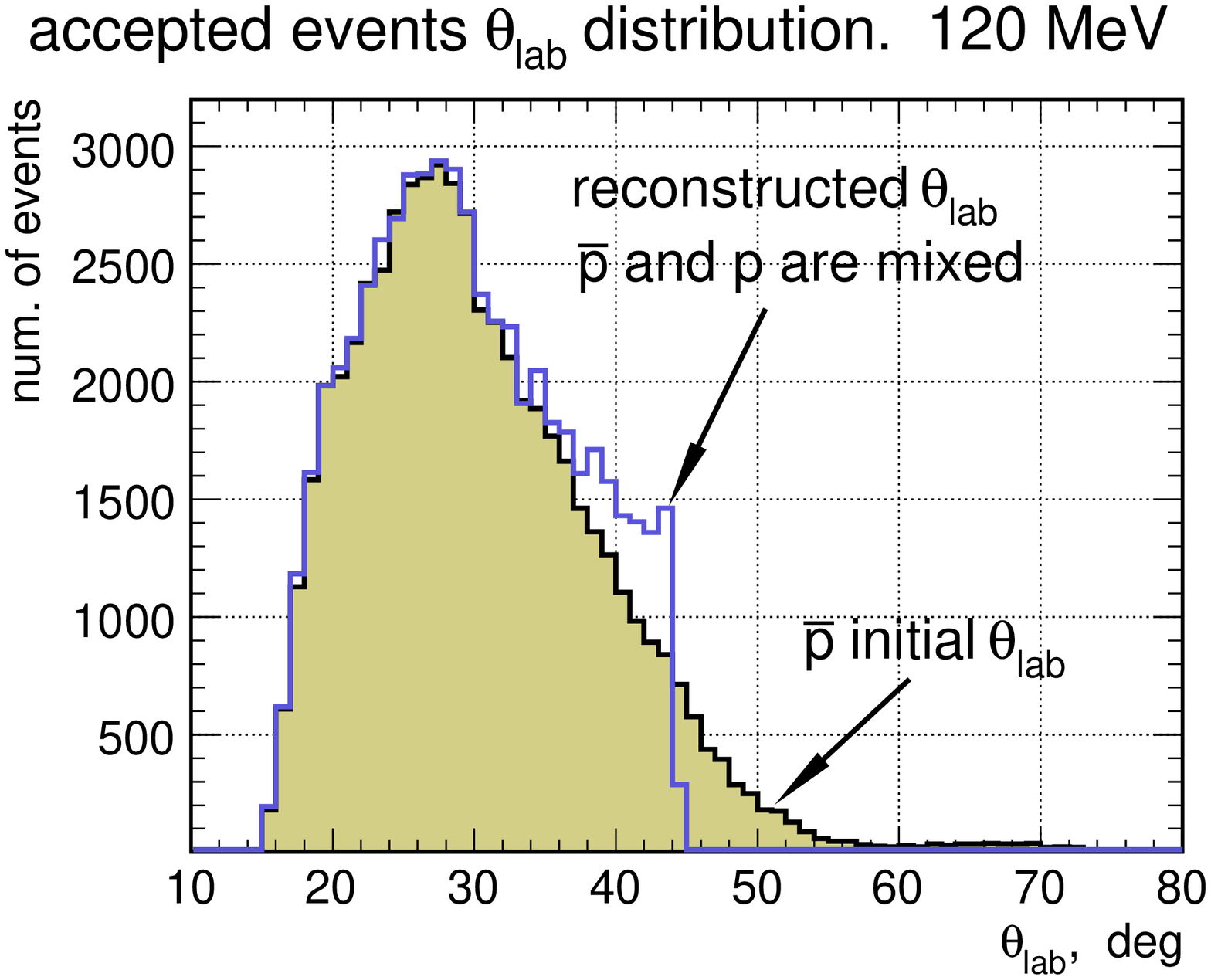}
 \parbox{14cm}{\caption{\label{fig:AcceptanceZandTheta}\small Left panel: Event distribution as a function of the vertex $z$ coordinate at 120~MeV for generated and reconstructed events. Right panel: Reconstructed event distribution as a function of the scattering angle $\vartheta_{lab}$ at 120~MeV. The filled histogram (yellow) shows the ideal case of truly identified antiprotons. The empty histogram (blue) shows the real case when all tracks with $\vartheta_{lab}<45^{o}$ are considered as scattered antiprotons. This leads to a small $\bar{p}p$ mixing effect at low beam energy (see text for details). It should ne noted that the distribution reflects also the non--uniformity of the geometrical acceptance  in the $\vartheta^{m}$ interval.}}
\end{center}
\end{figure}

An AD luminosity equal to $10^{27}$~cm$^{-2}$s$^{-1} = 1$~mb$^{-1}$s$^{-1}$ is assumed here at all energies. (During   spin--filtering for two beam lifetimes, the initial luminosity, given in Sec.~\ref{sec:beam-intensity}, has decreased by about a factor 7.)  The acceptance for elastic events is about 15\%, resulting in an event rate of the order of 10~s$^{-1}$, see Table~\ref{tab:Acceptance}. The background due to antiproton annihilations is estimated to be negligible (less than $10^{-4}$), since no background event survives out of $\rm 6\cdot 10^5$ primary annihilations.
In Fig.~\ref{fig:AcceptanceZandTheta}, the acceptance is shown as a function of the vertex coordinate ($z$, along the beam) at 120 MeV beam energy. The triangular distribution of all primary generated events corresponds to the density profile of the target gas.
\begin{table}[hbt]
\begin{center}
\begin{tabular}{|r|c|r|r|r|}
\hline
parameter & unit & $43\,MeV$ & $120\,MeV$ & $220\,MeV$  \\
\hline\hline
$\bar{p}p\to\bar{p}p$ -accepted&\% of $\bar{p}p\to all$     &  4.8  &  5.6 & 4.0 \\
                               &\% of $\bar{p}p\to\bar{p}p$ & 14.5  & 17.0 & 12.0 \\
\hline
reconstructed event rate & events/s & $12$ & $10$ & $5.8$ \\
'accepted' background events & \% & $0$ & $0$ & $0$ \\
\hline
\end{tabular}
 \parbox{14cm}{\caption{\label{tab:Acceptance}\small Acceptances, event rates and background contamination at the three considered beam energies: 43, 120, and 220 MeV. The AD luminosity is taken to be $10^{27}$~cm$^{-2}$s${^{-1}} = 1$~mb$^{-1}$s$^{-1}$.}} 
 \end{center}
\end{table}

Proton and antiproton electromagnetic interactions with matter are the same. Therefore, it is impossible to distinguish these two particles detected by silicon detectors if antiprotons do not annihilate. This leads to $\bar{p}p$ mixing: Events with antiproton scattering angle $\vartheta_{lab}>\pi/4$ are accepted through detection of forward protons. Thereby, the measured cross section and analyzing power will be degenerated.
On the other hand, the scattering cross section dominates in the forward hemisphere  which makes possible to measure the analyzing power (which is also slightly degenerated).

Introducing $\vartheta^{m}$ as {\it measured} scattering angle of a forward particle in the c.m.\ system $\vartheta^{m}=\vartheta$ if $\vartheta<\pi/2$, and $\vartheta^{m}=\pi - \vartheta$ if $\vartheta\ge\pi/2$, and the measured cross section 
\begin{equation}
	\label{eq:6}
\sigma^{m}_{o}(\vartheta^{m})\,=\,
\sigma_{o}(\vartheta^{m}) + \sigma_{o}(\pi-\vartheta^{m})~,
\end{equation} 
and the analyzing power
\begin{equation}
	\label{eq:7}
A_{y}^{m}(\vartheta^{m})\,=\,
\frac{\sigma_{o}(\vartheta^{m})}{\sigma_{o}(\vartheta^{m})+\sigma_{o}(\pi-\vartheta^{m})}A_{y}(\vartheta^{m})
+
\frac{\sigma_{o}(\pi-\vartheta^{m})}{\sigma_{o}(\vartheta^{m})+\sigma_{o}(\pi-\vartheta^{m})}A_{y}(\pi-\vartheta^{m})~,
\end{equation} 
we can deduce the beam polarization in the conventional way through
$A_{y}^{m}(\vartheta^{m})\,=\,\varepsilon^{m}(\vartheta^{m}) \cdot P$,
where $\varepsilon^{m}$ is the left--right asymmetry ($\sigma_{o}$ and $A_{y}$ are the corresponding true values depending on $\vartheta$, $P$ is the beam polarization).

\begin{figure}[hbt]
\begin{center}
\includegraphics[width=0.49\textwidth]{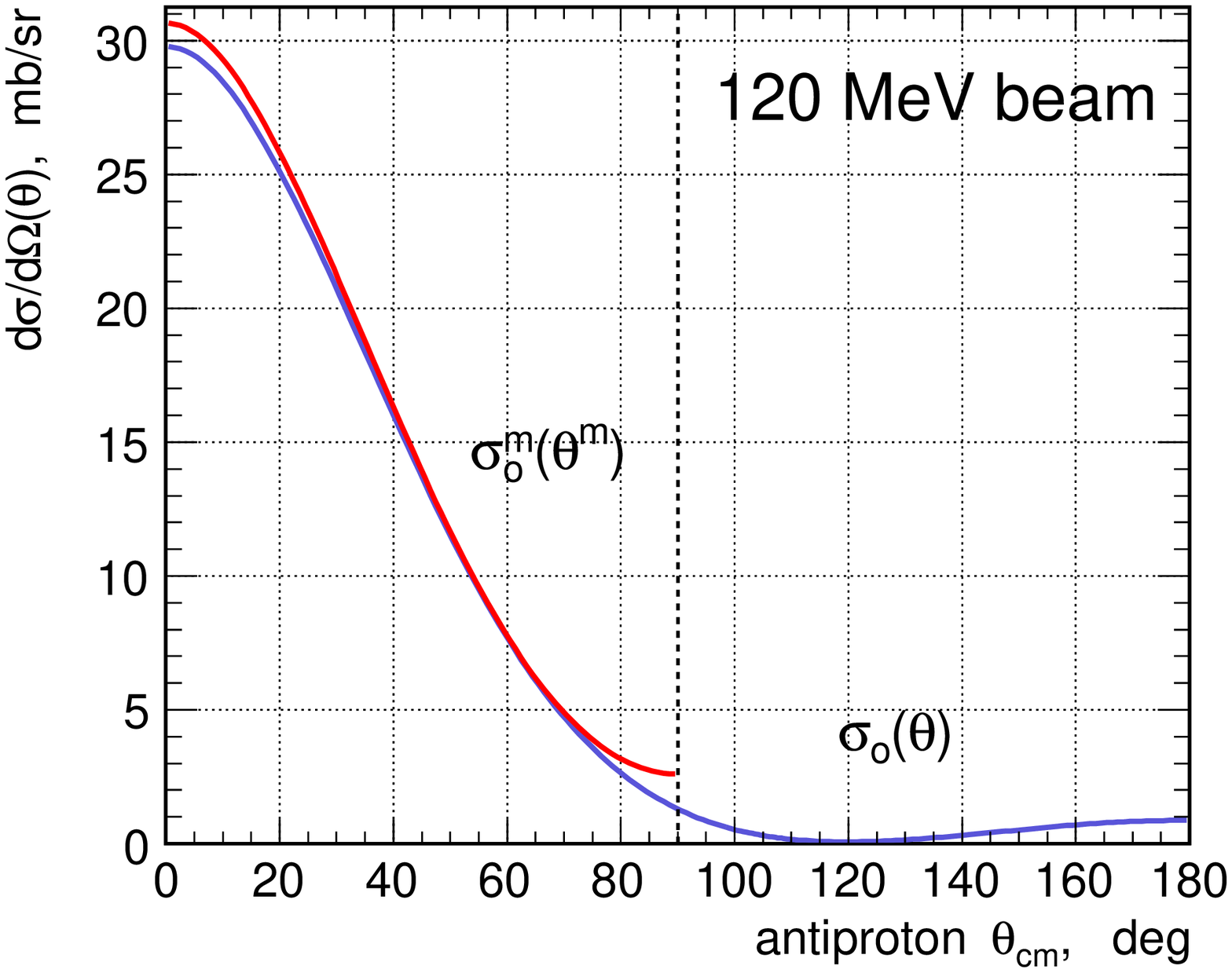}
\includegraphics[width=0.49\textwidth]{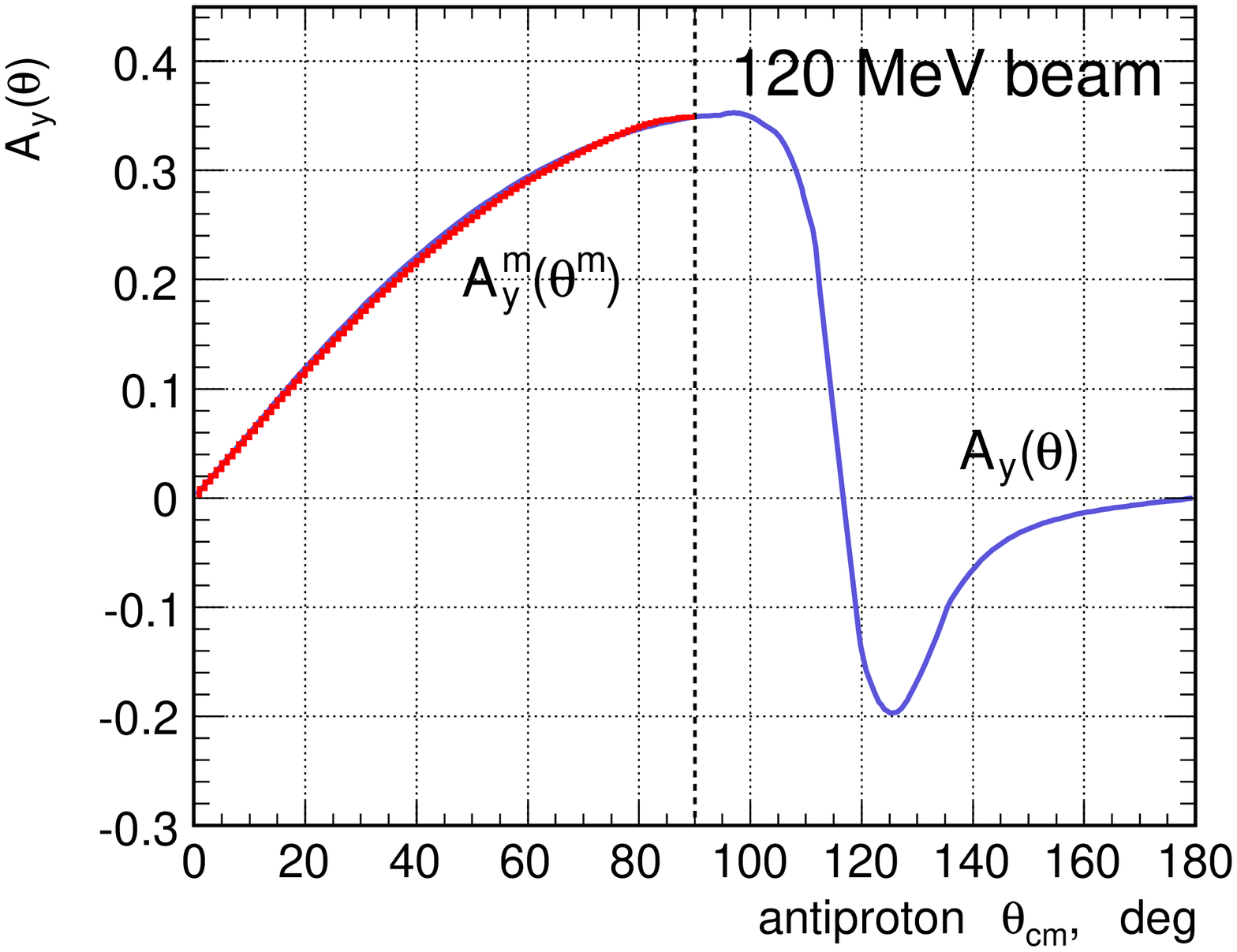}\\
 \parbox{14cm}{\caption{\label{fig:MixEff.CS}\small The model predictions~\cite{Mull,Hippchen,Haidenbauer} of $\sigma_{o}(\vartheta)$ and $A_{y}(\vartheta)$ are shown (in {\color{blue}blue}) with $\sigma_{o}^{m}(\vartheta^{m})$
and $A_{y}^{m}(\vartheta^{m})$ (in {\color{red}red)} for 120~MeV beam energy.}}  
\end{center}
\end{figure}

In Fig.~\ref{fig:MixEff.CS}, the model predictions of $\sigma_{o}(\vartheta)$ and $A_{y}(\vartheta)$ are shown (in blue) with $\sigma_{o}^{m}(\vartheta^{m})$ and $A_{y}^{m}(\vartheta^{m})$ (in red) for 120~MeV beam energy. Evidently, the mixing effect changes the measured cross sections and
analyzing powers only weakly.

Therefore, we can measure the antiproton beam polarization using the degenerated analyzing power values for the forward scattered particles. In any case, the existing experimental data on the $\bar{p}p$ elastic scattering cross section and the analyzing power allow us to to account correctly for the influence of the mixing effect.

Since multiple scattering dominates the resolution at the considered energies, almost no difference in the detector performance has been detected by reducing the strip pitch from 0.758~mm to 0.5~mm (see Table~\ref{tab:PitchDependence}). 

\begin{table}[hbt]
\begin{center}
\begin{tabular}{|r|r||c|c|c||c|c|c||} \hline
strip pitch & mm 
&
\multicolumn{3}{|c|}{\color{red} 0.5~mm }
&
\multicolumn{3}{|c|}{\color{blue} 0.758~mm }
\\ \hline\hline
beam energy & MeV 
& \color{red}43 & \color{red}120 & \color{red}220
& \color{blue}43 & \color{blue}120 & \color{blue}220
\\ \hline
$\sigma_{\vartheta_{lab}}$ & degree 
& \color{red}~0.77~ & \color{red}~0.35~ & \color{red}~0.25~ 
& \color{blue}~0.81~ & \color{blue}~0.42~ & \color{blue}~0.32~ 
\\
$\sigma_{\phi}$ & degree 
& \color{red}0.26 & \color{red}0.16 & \color{red}0.12 
& \color{blue}0.27 & \color{blue}0.18 & \color{blue}0.14 
\\
vertex $\sigma_{x}=\sigma_{y}$ & mm 
& \color{red}1.90 & \color{red}1.35 & \color{red}1.20 
& \color{blue}2.05 & \color{blue}1.47 & \color{blue}1.35 
\\
vertex $\sigma_{z}$ & mm 
& \color{red}0.34 & \color{red}0.19 & \color{red}0.14 
& \color{blue}0.35 & \color{blue}0.21 & \color{blue}0.17
\\
\hline
\end{tabular}
 \parbox{14cm}{\caption{\label{tab:PitchDependence}\small Reconstructed parameter uncertainties depending on strip pitch.}} 
\end{center}
\end{table}

We conclude that the same detector can be used for measurements at COSY ($pp$ elastic) and at the AD
($\bar{p}p$ elastic), since the reconstruction of elastic scattering in $pp$ and $\bar{p}p$ does not differ; the acceptance of the detector for elastic events is of the order of  15\%, resulting in an expected rate for reconstructed $\bar{p}p\to\bar{p}p$ elastic events around 10~s$^{-1}$; antiproton annihilation in the target environment as well as in the detector materials does not produce significant background and can be ignored; the detector resolution is mainly defined by multiple scattering, and a strip pitch of 0.76~mm is adequate; with increasing energy the acceptance slightly reduces (see Table~\ref{tab:Acceptance}), but $\bar{p}$ versus $p$ ambiguities are suppressed due to the dominating $\bar{p}$ scattering at small angles, at higher energies, there is a smaller annihilation rate and less track spread due to multiple scattering.

\cleardoublepage
\section{Timing of activities}
\pagestyle{myheadings} \markboth{Spin--Dependence
of the $\bar{p}p$ Interaction at the AD}{Timing of activities}
We suggest to organize the implementation of the experimental setup required for the measurements proposed here in a number of consecutive phases. This approach ensures that the regular AD operation for experiments using extracted beams is not adversely affected. An outline of this sequence of phases is given below, a detailed description follows in the subsequent sections.

\begin{enumerate}
 \item Commissioning of the low--$\beta$ section. Set up of the AD deceleration cycle with the new insertion.
 \item Commissioning of the experimental setup:
   \begin{itemize}
    \item[a)] Machine acceptance and stacking studies
    \item[b)] Measurement of the target polarization using antiprotons
   \end{itemize}
 \item Spin--filtering measurements with transverse polarization up to   $T_{\bar{p}}=70$~MeV 
       using the existing electron cooler.
 \item Upgrade of the electron cooler to 300~keV.  Set up of the AD deceleration cycle with upgraded electron cooler.
 \item Spin--filtering measurements with transverse polarization at $T_{\bar{p}}=50-450$~MeV.
 \item Implementation and commissioning of the Siberian snake.
 \item Spin--filtering measurements with longitudinal polarization at $T_{\bar{p}}=50 - 450$~MeV.
\end{enumerate}

It should be noted that it is important for the collaboration that we obtain a go--ahead from the SPS committee for the first three phases in order to set the chain of actions into motion as soon as possible. We would like to emphasize that the funds available to us suffice to carry out the measurements in phases~1--3. We are currently acquiring the additional funding needed for the upgrade of the electron cooler and the construction of the Siberian snake. 

\subsection{Phases of the PAX experimental program at the AD}

\subsubsection{Phase 1: Commissioning of low--$\beta$ section}
\begin{figure}[t]
 \begin{center}
\includegraphics[width=0.7\linewidth]{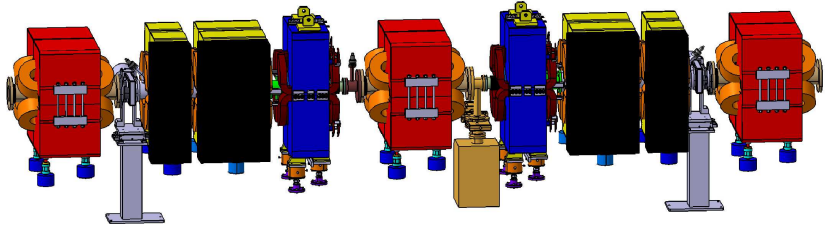}
  \parbox{14cm}{\caption{\label{fig:setup-phase1}\small After the installation of two gate valves, six additional quadrupole magnets will be installed in phase~1. During the commissioning of the low--$\beta$ section, the central quadrupole magnet QDN15 will remain in place.
}}
\end{center}
\end{figure}

In phase~1, six quadrupole magnets will be installed in the straight section at 3 o'clock (see Fig.~\ref{fig:AD-layout}) to provide the required small $\beta$--function to operate the polarized PIT. The lattice calculations have been already reported in Sec.~\ref{sec:low-beta}. To satisfy the requirements from the calculations, two  quadrupoles of the former CELSIUS ring in Uppsala and four spare quadrupoles of the  COSY ring at J\"ulich are available to be installed at the AD. In order to allow for the installation of the quadrupole magnets without negatively affecting the beamline vacuum of the AD ring, as a first step, two gate valves shall be installed. The foreseen complete setup of phase~1 is depicted in Fig.~\ref{fig:setup-phase1}.

During phase 1, the central quadrupole of the AD section (QDN15) will stay in place and dedicated machine studies will be devoted to the commissioning of the low--$\beta$ section. It is anticipated that the commissioning work can be carried out parasitically during 2010 without devoting dedicated machine development time at the AD for this task. Together with the quadrupole magnets, a set of four steerer magnets (2$\times$horizontal and 2$\times$vertical) and two Beam Position Monitors (BPM) will be installed to control the machine orbit in the low--$\beta$ insertion. Because of spacial restrictions the steerer windings have to be mounted on the yokes of the quadrupole magnets. The machine development shall ensure that the QDN15 can effectively be replaced by the combined function of the new quadrupoles and that all regular AD operations can be accomplished without this magnet. 

After the new magnets have been successfully commissioned, providing the usual AD operation, additional machine development is necessary to operate the new quadrupole magnets with minimal $\beta$--functions for the PAX experiment. The installation of NEG coated beam tubes inside the additional six quadrupole magnets of the low--$\beta$ insertion shall be carried out at the same time when the magnets are installed, because  additional pumping capacity is needed. Ideally, after the machine development is accomplished, it should be possible to switch the AD optics back and forth on short notice from the regular machine optics, providing extracted beams, to the machine operation required  by PAX. 

If the new six quadrupole magnets would be installed in the coming winter shutdown 2009--2010, we estimate that one year of parasitic machine development would be sufficient. The machine could then be ready at the end of 2010 for the new tasks involved in the measurements proposed here.

\subsubsection{Phase 2: Commissioning of the experimental setup}
\begin{figure}[t]
 \begin{center}
\includegraphics[width=0.70\linewidth]{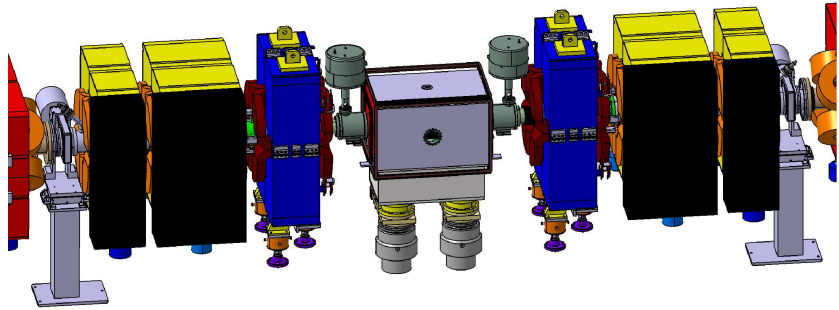}
  \parbox{14cm}{\caption{\label{fig:target-chamber-only}\small In phase~2, the target chamber replaces the central quadrupole magnet QDN15. Two fast gate valves are mounted up-- and downstream of the target chamber.
}}
\end{center}
\end{figure}

Once phase~1 is successfully implemented, the central quadrupole magnet QDN15 can be removed without risk in the winter shutdown 2010/11  and the target chamber can be installed. The experimental setup available at this point is shown in Fig.~\ref{fig:target-chamber-only}. 

Phase~2 is composed of two parts. During the first part, in  phase~2a, once the target chamber is installed at the AD, two important machine aspects related to the  measurements proposed here can already be addressed without the PIT  installed at the AD. The second part, phase~2b, addresses first measurements using the polarized target.

\begin{itemize}
 \item {\bf Phase 2a: Machine acceptance and stacking studies} 
    \begin{enumerate}
      \item It is important to determine the machine acceptance with electron cooled beam. The      technique we have established for this purpose is described in ref.~\cite{grigoryev}. It uses a small movable frame to scrape the beam.  The advantage over other methods is that it also yields a measurement of the machine acceptance angle at the location of the target. This parameter needs to be precisely measured because it plays a crucial role in the interpretation of the observed polarization buildup, as discussed in Sec.~\ref{sec:buildup}. We estimate that these machine acceptance studies can be accomplished with two weeks of dedicated machine development time.
      \item The proposed measurements  would greatly benefit from higher luminosities, as already discussed in Sec.~\ref{sec:beam-intensity}. Therefore, we suggest to explore at an early stage what could be achieved through stacking of antiprotons into the AD.  We estimate that the stacking studies can be accomplished with two weeks of dedicated machine development time.
   \end{enumerate}
\end{itemize}

The installation of the polarized internal target, the complete detector system, and the BRP are foreseen to take place in the winter shutdown 2011/2012. 
\begin{itemize}
 \item {\bf Phase 2b: Measurement of the target polarization using antiprotons}\\ These first measurements will address the determination of the target polarizations when H is injected into the storage cell. We anticipate that three weeks will be sufficient to accomplish this goal.
\end{itemize}

\subsubsection{Phase 3: Spin--filtering measurements with transverse polarization up to  $T_{\bar{p}}=70$~MeV using the existing electron cooler}
\begin{figure}[t]
 \begin{center}
\includegraphics[width=0.7\linewidth]{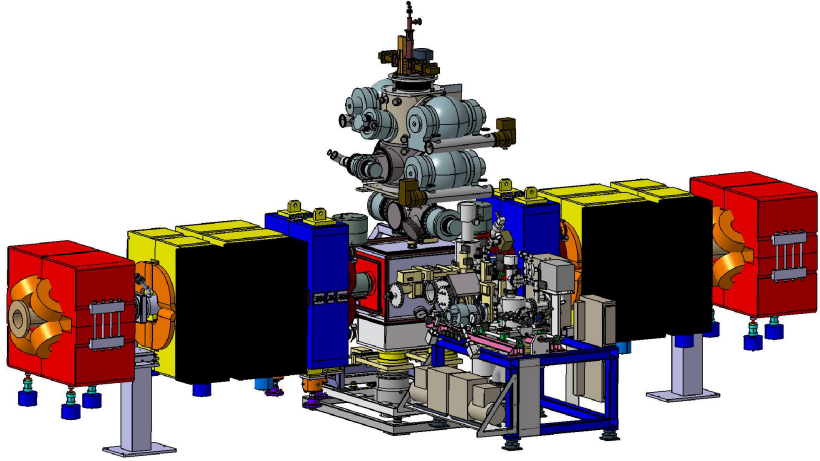}
  \parbox{14cm}{\caption{\label{fig:setup-overview2}\small Full installation foreseen in phase~3 at the AD. The atomic beam source is mounted above the target chamber that houses the detector system and the storage cell. The Breit--Rabi target polarimeter and the target--gas analyzer are mounted outwards of the ring.}}
\end{center}
\end{figure}

After phase~2, the necessary requirements for the first polarization buildup studies at the AD are fulfilled.  The setup available at this point is shown in Fig.~\ref{fig:setup-overview2} (it was already shown in Fig.~\ref{fig:setup-overview}).

Phase~3 will be carried out using transversely polarized beam produced by spin--filtering using a transversely polarized H or D target. These studies can be carried out without any modifications at the AD electron cooler at energies between 30 and 70~MeV. As shown in Fig.~\ref{fig:predictedPol}, spin--filtering at lower energies, below 30~MeV, is not very effective because Rutherford scattering  is dominating the loss of beam particles. Unfortunately, the existing AD electron cooler does not permit to provide electron cooled stored beam above about 70~MeV. It is anticipated that six weeks of beamtime are sufficient to accomplish phase~3.

\subsubsection{Phase 4: Upgrade of the electron cooler to 300~keV}
The filtering experiments require to compensate multiple scattering in the target by electron cooling. The present AD electron cooler is capable to provide electron energies of up to 30~keV, corresponding to antiproton beam energies of 70~MeV. In order to carry out the proposed measurements in the energy range between 50 and 450 MeV, the electron cooler at the AD should be upgraded to about 300~keV. The upgrade of the electron cooler is anticipated to be carried out during the winter shutdown 2012/2013. We anticipate that for the commissioning of the electron cooler with beam about four weeks of machine development are necessary in 2013.

\subsubsection{Phase 5: Spin--filtering measurements with transverse polarization at $T_{\bar{p}}=50-450$~MeV}
Once the energy range for the spin--filtering studies at the AD has been expanded through the upgrade of the electron cooler up to beam energies of about 450~MeV, six week of beam time in 2013 should be sufficient for the measurements using H and D targets.

\subsubsection{Phase 6: Implementation and commissioning of the Siberian snake}
In order to determine $\sigma_2$, the stable beam spin direction has to be longitudinal at the position of the PIT. Therefore, in the straight section opposite the PIT, a solenoidal Siberian snake must be implemented (see Fig.~\ref{fig:AD-layout}). We have begun to investigate whether existing snakes can be utilized or modified to be used at the AD. In any case, a careful machine study has to be carried out before final conclusions can be reached. We anticipate that such a snake could be available for installation into the AD during the winter shutdown 2013/2014.

\subsubsection{Phase 7: Spin--filtering measurements with longitudinal polarization at $T_{\bar{p}}=50 - 450$~MeV}
With a Siberian snake, polarization buildup studies with longitudinally polarized target shall be carried out in 2014. We anticipate that this phase requires about eight weeks of beam time.

\subsection{Anticipated time plan}
Below, we give an approximate timetable for the activities outlined in the present proposal.
\begin{center}
\begin{tabular}{l|p{3.3cm}|p{9.9cm}}
Phase  		& Time			& Description \\\hline
	1 	& Shutdown 2009/10	& Installation of six magnets for the low--$\beta$ insertion, the central quadrupole remains in place.\\
                & 2010			& Commissioning of the low--$\beta$ section, carried out parasitically during (possibly extended) machine development sessions.\\ \hline
2a		& Shutdown 2010/11	& Installation of the target chamber.\\
		& 2 weeks in 2011	& Machine acceptance studies.\\
		& 3 weeks in 2011	& Stacking studies (split in sessions). \\
		&			&		   \\
2b		& Shutdown 2011/12	& Installation of ABS, BRP, and detector system.\\
		& 2 weeks in 2012	& Measurement of H target polarization with antiproton beam. \\\hline
3		& 6 weeks in 2012	& Spin--filtering measurements using  H and D target up to 70~MeV with transverse beam polarization. \\\hline
4		& Shutdown 2012/13	& Shutdown possibly extended up to six months. Implementation and commissioning of electron cooler upgrade to 300~keV.\\\hline
5		& 4 weeks in 2013	& Commissioning of the electron cooler with beam\\
		& 6 weeks in 2013	& Spin--filtering measurements using H and D target up to 430~MeV with transverse beam polarization. \\\hline
6		& Shutdown 2013/14	& Implementation of the Siberian snake. \\\hline
7		& 8 weeks in 2014	& Spin--filtering measurements with H and D target up to 430~MeV with longitudinal beam polarization, including commissioning of the Siberian snake. \\\hline
\end{tabular}
\end{center}

%
%

\cleardoublepage
\pagestyle{myheadings} \markboth{Spin--Dependence
of the $\bar{p}p$ Interaction at the AD}{Acknowledgment}
\section{Acknowledgment}
We would like to thank the members of the AD machine crew, Maria Elena Angoletta, Pavel Belochitskii, Fritz Caspers, Stephan Maury, Dieter M\"ohl, Flemming Pedersen, Gerard Tranquille, and Tommy Eriksson for their support and the many helpful suggestions brought up during our site visits.

\pagestyle{myheadings} \markboth{Spin--Dependence of the
$\bar{p}p$ Interaction at the AD}{Bibliography}

\end{document}